\documentclass[english,american]{myclass}
\usepackage[T1]{fontenc}
\usepackage{babel}
\usepackage{prettyref}
\usepackage{units}
\usepackage{amstext}
\usepackage{amssymb}
\usepackage{graphicx}
\usepackage{esint}
\usepackage[authoryear]{natbib}
\usepackage[unicode=true]
 {hyperref}
\usepackage{breakurl}

\makeatletter

\providecommand{\tabularnewline}{\\}

\usepackage{xcolor}
\definecolor{darkblue}{rgb}{0.0,0.0,0.64}

\definecolor{choose_color}{HTML}{B0CFEC}
\definecolor{connect_color}{HTML}{EDAF9B}
\definecolor{rest_color}{HTML}{EBD7A0}

\usepackage{listings}
\lstdefinestyle{C++}{morekeywords={size_t}}
\usepackage{lstlinebgrd}

\usepackage{amsmath}

\usepackage{prettyref}
\newrefformat{chap}{\hyperref[#1]{Chapter~\ref*{#1}}}
\newrefformat{sec}{\hyperref[#1]{Section~\ref*{#1}}}
\newrefformat{sub}{\hyperref[#1]{Section~\ref*{#1}}}
\newrefformat{apdx}{\hyperref[#1]{Appendix~\ref*{#1}}}
\newrefformat{lst}{\hyperref[#1]{Listing~\ref*{#1}}}
\newrefformat{fig}{\hyperref[#1]{Figure~\ref*{#1}}}
\newrefformat{tab}{\hyperref[#1]{Table~\ref*{#1}}}
\newrefformat{eqn}{\hyperref[#1]{Equation~\ref*{#1}}}

\usepackage{url,hyperref,lineno,microtype,subcaption}
\usepackage[onehalfspacing]{setspace}

\hypersetup{colorlinks=true, breaklinks=true, linkcolor=darkblue, urlcolor=darkblue, citecolor=darkblue}

\usepackage{soul}
\usepackage{tabularx}
\usepackage{color}
\usepackage{colortbl}
\usepackage{morefloats}
\usepackage{placeins}

\usepackage[font={scriptsize}]{caption}

\def\keyFont{\fontsize{8}{11}\helveticabold }
\def\firstAuthorLast{Senk {et~al.}} 
\def\Authors{Johanna Senk\,$^{1,*}$, Corto Carde\,$^{2,3,4}$, Espen Hagen\,$^{1,5}$, Torsten W. Kuhlen\,$^{2,3}$, Markus Diesmann\,$^{1,6,7}$ and Benjamin Weyers\,$^{2,3}$}

\def\Address{
$^{1}$Institute of Neuroscience and Medicine (INM-6) and Institute for Advanced Simulation (IAS-6) and JARA Institute Brain Structure-Function Relationships (INM-10), J\"{u}lich Research Centre, J\"{u}lich, Germany\\
$^{2}$Visual Computing Institute, RWTH Aachen University, Aachen, Germany\\
$^{3}$JARA - High-Performance Computing, Aachen and J\"{u}lich, Germany\\
$^{4}$IMT Atlantique Bretagne-Pays de la Loire, Brest, France\\
$^{5}$Department of Physics, University of Oslo, Oslo, Norway\\
$^{6}$Department of Psychiatry, Psychotherapy and Psychosomatics, Medical Faculty, RWTH Aachen University, Aachen, Germany\\
$^{7}$Department of Physics, Faculty 1, RWTH Aachen University, Aachen, Germany}

\def\corrAuthor{Johanna Senk}
\def\corrEmail{j.senk@fz-juelich.de}

\makeatother

\begin{document}
\onecolumn
\firstpage{1}
\author[\firstAuthorLast ]{\Authors}  
\address{}                            
\correspondance{}                     
\extraAuth{}                          
\title[VIOLA]{VIOLA - A multi-purpose and web-based visualization tool for neuronal-network simulation output} 
\phantomsection\addcontentsline{toc}{section}{Title}
\maketitle

\begin{abstract}
Neuronal network models and corresponding computer simulations are
invaluable tools to aid the interpretation of the relationship between
neuron properties, connectivity and measured activity in cortical
tissue. Spatiotemporal patterns of activity propagating across the
cortical surface as observed experimentally can for example be described
by neuronal network models with layered geometry and distance-dependent
connectivity. In order to cover the surface area captured by today's
experimental techniques and to achieve sufficient self-consistency,
such models contain millions of nerve cells. The interpretation of
the resulting stream of multi-modal and multi-dimensional simulation
data calls for integrating interactive visualization steps into existing
simulation-analysis workflows. Here, we present a set of interactive
visualization concepts called views for the visual analysis of activity
data in topological network models, and a corresponding reference
implementation VIOLA (VIsualization Of Layer Activity). The software
is a lightweight, open-source, web-based and platform-independent
application combining and adapting modern interactive visualization
paradigms, such as coordinated multiple views, for massively parallel
neurophysiological data. For a use-case demonstration we consider
spiking activity data of a two-population, layered point-neuron network
model incorporating distance-dependent connectivity subject to a spatially
confined excitation originating from an external population. With
the multiple coordinated views, an explorative and qualitative assessment
of the spatiotemporal features of neuronal activity can be performed
upfront of a detailed quantitative data analysis of specific aspects
of the data. Interactive multi-view analysis therefore assists existing
data analysis workflows. Furthermore, ongoing efforts including the
European Human Brain Project aim at providing online user portals
for integrated model development, simulation, analysis and provenance
tracking, wherein interactive visual analysis tools are one component.
Browser-compatible, web-technology based solutions are therefore required.
Within this scope, with VIOLA we provide a first prototype.

\keyFont{\section{Keywords:}interactive visualization, visual data
analytics, coordinated multiple views, 3D visualization, neuronal
network simulation, spiking neurons, spatiotemporal patterns, data
analysis workflow}
\end{abstract}

\section{Introduction}

\label{sec:introduction}

One common technique to capture brain activity on the neuronal level
is to record extracellular potentials in cortical tissue \citep{Buzsaki2012,Einevoll2013}.
The low frequency ($\unit[\lesssim100]{Hz}$) part of the signal,
often referred to as the local field potential (LFP), remains difficult
to interpret as thousands to millions of proximal and distal neurons
contribute to the signal \citep{Kajikawa2011,Linden2011,Leski2013}.
From the high-frequency band ($\unit[\gtrsim100]{Hz}$), however,
one can detect sequences of spikes, the transient extracellular signatures
of action potentials in single neurons nearby the recording electrode.
The number of reliably identified neurons (through spike sorting,
\citealp{Quiroga2007}) per recording session is low compared to the
number of neurons in vicinity of the recording device, even if the
experiment is performed with hundreds or more electrode contact points
\citep{Einevoll2012}. The Utah array from Blackrock Microsystems\footnote{\href{http://blackrockmicro.com}{http://blackrockmicro.com}},
for example, resolves with $10\times10$ electrodes on $\unit[4\times4]{mm^{2}}$
little more than a hundred distinct neurons. Also optical methods
for measuring neuronal activity have seen continuous improvements.
As recently demonstrated, non-invasive three-photon fluorescence microscopy
facilitates functional imaging at high optical resolution as deep
as $\unit[1]{mm}$ \citep{Ouzounov2017}. While the method simultaneously
images a comparably large number of neurons, the recordings lack the
temporal resolution to reliably detect individual action potentials.
\citet{Ouzounov2017} record from as many as $150$ neurons in mouse
hippocampal stratum pyramidale within a field of view of $\unit[200\times200]{\mu m}$. 

The rapidly improving parallel recording technology increases the
need for suitable analysis methods for high-dimensional and dynamic
data streams. Nevertheless, the recordings will remain to be characterized
by a massive undersampling for some time. Therefore, detailed full
scale models of the cortical tissue are required to understand the
microscopic dynamics \citep{Albada15} and to relate the microscopic
activity to mesoscopic measures like the LFP. For this program to
succeed, neuroscientists not only need to analyze model data in the
same way as experimental data, but to explore data sets with orders
of magnitude more channels than experimentally available. 

Networks of model neurons incorporating varying levels of biophysical
and anatomical detail reproduce a number of features of experimentally
obtained spike trains. For networks of point- or one-compartment neuron
models, this list of features includes irregular spike trains \citep{Softky1993,Vreeswijk1996,Amit1997,Shadlen1998},
asynchronous spiking \citep{Ecker2010,Renart2010,Helias14,Ostojic2014},
correlation structure \citep{Gentet2010,Okun2008,Helias2013}, self-sustained
activity \citep{Ohbayashi2003,Kriener2014}, realistic firing rates
across cortical lamina \citep{Potjans2014}, single-neuron spiking
activity of different cell types \citep{Izhikevich2003,Kobayashi2009,Yamauchi2011}
and responses under \textit{\emph{\textquoteleft in vivo\textquoteright{}}}
conditions \citep{Jolivet2008,Gerstner2009}. Relating point-neuron
network activity to population signals such as the LFP is, however,
not straightforward. Approximations \citep[see][]{Mazzoni2015} or
forward-model based schemes \citep{Hagen2016} are required to bridge
the gap to experimental electrophysiological data which predominantly
reflects population activity.

The focus of this study lies on visualization methods for activity
of spatially extended neuronal network models. Incorporation of spatial
structure is a prerequisite for models aiming to explain experimentally
observed spatiotemporal patterns of activity \citep{Rubino2006,Denker2011,Sato2012,Muller2014,Townsend2015}.
Such models have an arrangement of neurons in one-, two- or three-dimensional
(1D, 2D or 3D) space and connection rules which typically depend on
the distance between (parts of) the neurons \citep{Mehring2003,Coombes2005,Yger2011,Bressloff2012,Voges2012,Kriener2014,Keane2015,Rosenbaum2017}.
Although we primarily focus on model data, the same visualization
methods can be applied with experimentally recorded data. 

\begin{figure}[t]
\includegraphics[width=1\textwidth]{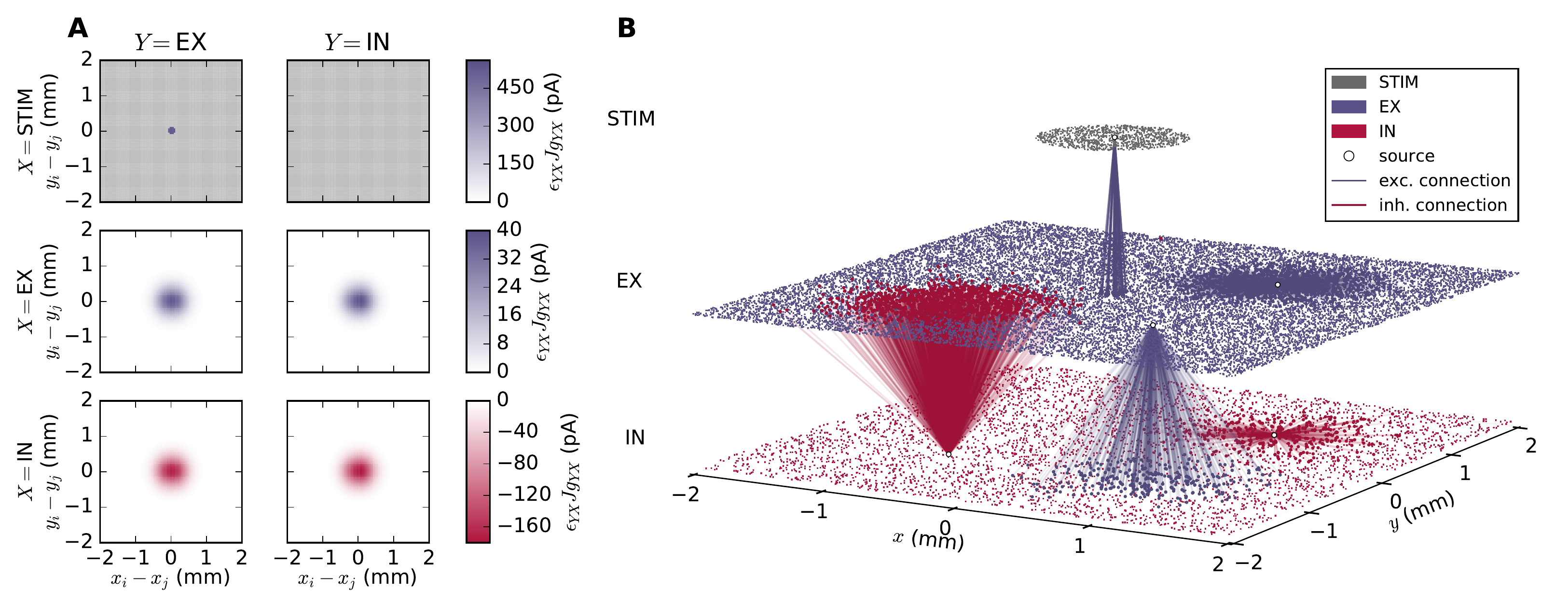}
\caption{\textbf{Geometry and connectivity of a layered point-neuron network.}
\textbf{A}~Schematic illustration of distance-dependent network connectivity
using connectivity pattern tables \citep{Nordlie2010}. Each row represents
source populations $X\in\{\mathrm{STIM,EX,IN}\}$, and each column
target populations $Y\in\{\mathrm{EX,IN}\}$. The color coding in
each image shows the connection intensity between presynaptic neurons
$j$ and postsynaptic neurons $i$ located in $(x_{j},y_{j})$ and
$(x_{i},y_{i})$ with origin $\left(0,0\right)$ at the center. The
connection intensities are defined as the product between pairwise
connection probabilities $\epsilon_{YX}(r_{ij})$ and synapse strengths
$g_{YX}J$ for each respective connection. Gray values denote connection
intensities of zero. \textbf{B}~Illustration of one network instantiation
with randomly drawn neuron positions and outgoing connections from
a subset of neuronal units. The colored dots represent individual
units at their $(x,y)$-coordinates. Gray dots denote units in a stimulus
($\mathrm{STIM})$ layer, blue dots excitatory ($\mathrm{EX}$) units,
and red dots inhibitory ($\mathrm{IN}$) units. Blue and red lines
denote excitatory and inhibitory connections respectively, from a
source unit (white circles) onto neurons within the same or another
layer. }
\label{fig:network_sketch} 
\end{figure}

\begin{figure}[t]
\includegraphics[width=1\textwidth]{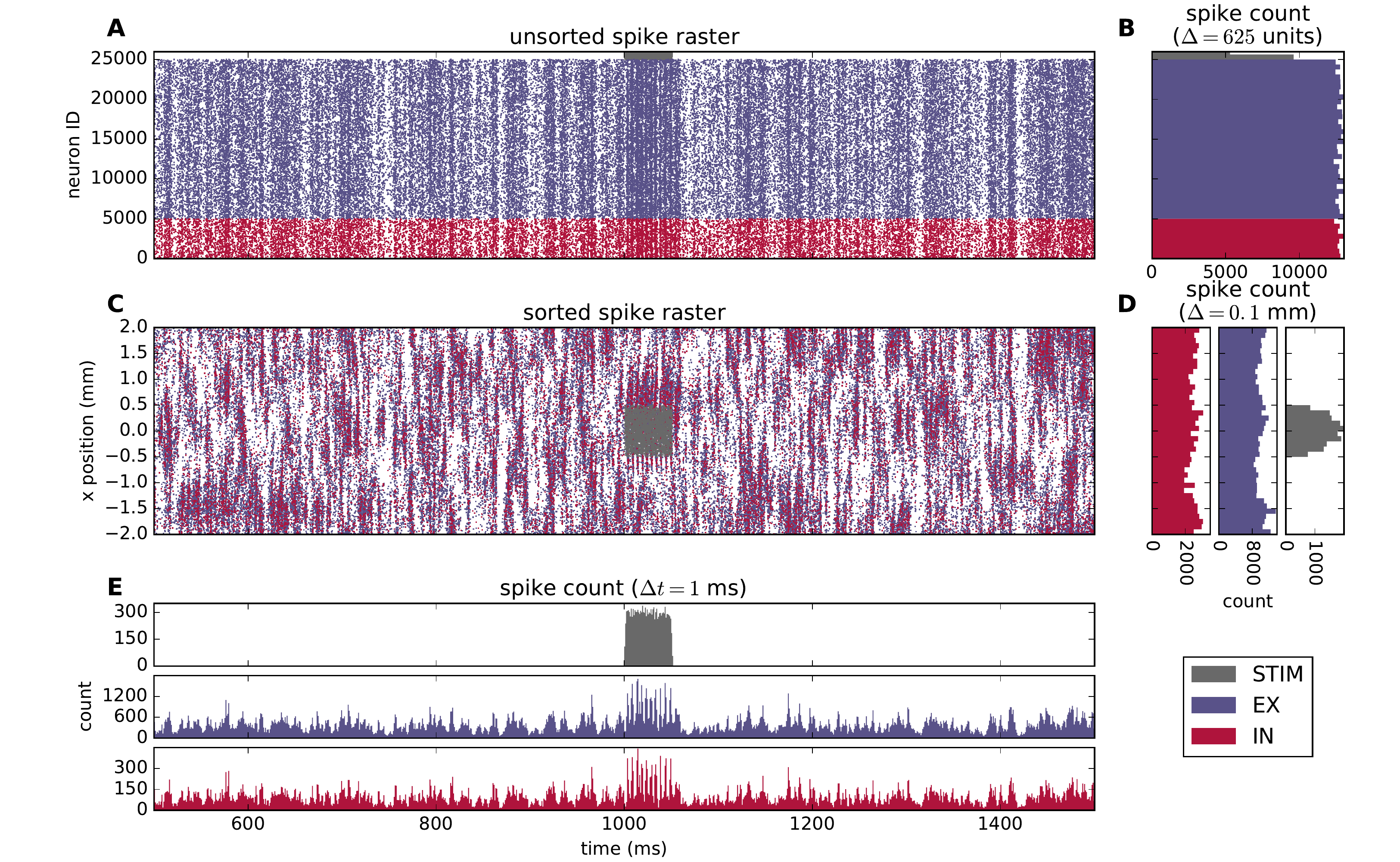} \caption{\textbf{Spiking activity of a layered point-neuron network model.}
\textbf{A}~Spike raster plot for $\mathrm{STIM}$ (gray dots), $\mathrm{EX}$
(blue dots) and $\mathrm{IN}$ (red dots) units from a simulation
of the network instantiation depicted in \prettyref{fig:network_sketch}B.
Each individual dot corresponds to a unit ID vs. spike time; only
the spikes of every fifth neuron are shown in the raster. The color
coding for each population is reused in the subsequent panels. \textbf{B}~Spike
count histogram across units in each population, calculated using
a bin width of $625$ units, sorted by neuron index $j$. \textbf{C}~Sorted
spike raster where dots correspond to the spatial location (projected
onto the $x-$axis) and spike times of each unit. The raster-plot
density is diluted as in Panel~A. \textbf{D}~Spike count histogram
across spatial bins with a width $\Delta l=\unit[0.1]{mm}$. \textbf{E}~Spike
count histogram for each population across time, computed using a
temporal bin width $\Delta t=\unit[1]{ms}$. }
\label{fig:raster} 
\end{figure}

We here consider an example spiking point-neuron network consisting
of an excitatory ($\mathrm{EX})$, an inhibitory ($\mathrm{IN}$)
and an external stimulus ($\mathrm{STIM}$) population. $\mathrm{EX}$
and $\mathrm{IN}$ units are positioned randomly within square domains
while $\mathrm{STIM}$ units are randomly positioned within a circle
at the center. A schematic representation of the network connectivity
is shown in \prettyref{fig:network_sketch}A. We use connectivity
pattern tables \citep{Nordlie2010} for source populations $X$ (rows)
and target populations $Y$ (columns). The images indicate the \textquoteleft connection
intensities\textquoteright{} for each connection, defined as the product
between averaged pairwise connection probabilities $\epsilon_{YX}(r_{ij})$
and synapse strengths $g_{YX}J$. The distance between a source and
a target neuron is denoted by $r_{ij}$. Pairwise connection probabilities
decay with horizontal distance between $\mathrm{EX}$ and $\mathrm{IN}$
units according to a Gaussian profile, while $\mathrm{STIM}$ units
only connect locally to the $\mathrm{EX}$ population restricted by
a cut-off radius. The geometry of one network instantiation is depicted
in \prettyref{fig:network_sketch}B. $\mathrm{EX}$ (blue dots), $\mathrm{IN}$
(red dots) and $\mathrm{STIM}$ (gray dots) units are placed in separate
layers. The distance dependency is illustrated by outgoing excitatory
connections (blue lines) from single units in the $\mathrm{STIM}$
and $\mathrm{EX}$ populations and outgoing inhibitory connections
(red lines) from single units in the $\mathrm{IN}$ population.

The visualization of neuronal activity data poses challenges due
to the high dimensionality and time dependence of the data. Historically,
electrophysiological data have been recorded from few electrodes or
from many electrodes with undefined relative and absolute spatial
coordinates \citep[see, for example, the pioneering work of][]{Krueger81_191}.
This is not an essential limitation for recordings within the local
cortical network where a neuron can form a synapse with any other
neuron and there is little spatial organization. Furthermore, the
fundamental interaction in a neuronal network is considered to be
a dynamics on a graph; nodes solely interact via the edges of the
graph. In this picture the spatial embedding of the graph is arbitrary
as the dynamics is not constrained by the dimensions of physical space.
Therefore, neuroscientists developed concepts for visualizing correlation
structure and time dependence of neuronal activity in multi-channel
recordings in ignorance of spatial properties. A temporal segment
of activity of our example network is visualized in \prettyref{fig:raster}.
Panel A is the spike raster diagram or dot display in use for decades
\citep[explained in][]{Abeles82}. Each row shows the spike train
of one neuron where spike times are marked by dots. The rows either
represent data of the same neuron in several trials or, as here, data
of simultaneously recorded neurons in a single trial \citep[Fig. 6.2]{GruenPhD}.
The spike trains are vertically arranged by neuron ID and in addition
color coded by population. The spike raster highlights global features
of network activity and generations of neuroscientists have been trained
to visually inspect these diagrams. For example, network synchrony
appears as a stripy vertical pattern even if individual neurons only
rarely participate in an individual synchronous event. The activation
of the stimulus population is reflected in the other populations as
an increased density of the dots. \citet{Epping1984} extend the concept
of the raster diagram by assigning a unique color to the dots of a
neuron. In this way multi-channel activity observed over multiple
trials can be superimposed. Panel B shows spike counts along the temporal
axis over neuronal units demonstrating that the per-neuron spike count
is similar for the excitatory and the inhibitory populations. The
spike count along the vertical axis in panel E is called the post-stimulus
time histogram \citep[PSTH,][]{Perkel67a} originally computed for
an individual neuron observed over several trials. Later the display
was also called peri-stimulus time histogram. Here the histogram is
computed over simultaneously recorded neurons in a single trial. The
display uncovers the fluctuations of population activity in time.

The development of adequate visualization concepts for multi-channel
neuronal data is an ongoing endeavor \citep{Allen2012}. The cross-correlation
function \citep{Perkel67b} exposes the time-averaged relationship
between the spike times of two neurons. The snowflake diagram generalizes
the concept to three neurons \citep{Perkel75,Czanner05_1456}. Gravitational
clustering (\citealp{Gerstein85b,Gerstein85a} reviewed in Chap. 8
of \citealp{GruenRotter10}) attempts to identify the emergence of
correlated groups of neurons, so called cell-assemblies, and the temporal
dynamics of the changing membership of individual neurons in such
groups without averaging over trials. The joint peri-stimulus time
histogram \citep[JPSTH, ][]{Aertsen89} generalizes the cross-correlation
function to visualize the dynamics of the correlation between the
spikes of two neurons in reference to a stimulus. Later \citet{Prut98}
used the idea to investigate the occurrence of spatiotemporal patterns
in the spike trains of three neurons, where ``spatio'' refers to
the abstract space of neuron IDs not physical space. Because of the
difficulties in determining statistical significance \citet{Gruen02_81}
restricted the scope to patterns in the space of IDs and for visualization
mapped significant events, so called unitary events, back into the
spike raster diagram. Progress in the theory of neuronal networks
showed that propagating spiking activity due to the stochastic nature
of neuronal activity is likely to exhibit in each instance a random
sub-pattern of spikes of some superset of neurons. Therefore \citet{Schrader08}
designed a matrix spanned by binned ongoing time in both dimensions
where matrix-elements represent the cardinality of the intersection
set of the neurons spiking at the two respective time bins. With color-coded
cardinality, in this matrix repeatedly occurring propagating spiking
activity appears as a diagonal feature. Later an assessment of statistical
significance was added \citep{Torre16_e1004939}. \citet{Kemere2008}
employ multi-channel recordings to construct the time course of a
multi-dimensional vector of spike rates. A suitable projection to
a lower dimensional space reveals differentiable trajectories of network
activity depending on the experimental protocol \citep[reviewed in][]{Cunningham2014}.
Another line of work attempts to cope with the combinatorial explosion
of patterns in multi-channel spike trains while maintaining sensitivity
by the construction of a pattern spectrum: a two-dimensional histogram
spanned by the number of spikes in a pattern, called pattern complexity,
and the number of occurrences of the particular patterns \citep{Gerstein12_54,Torre2013}. 

\prettyref{fig:raster}C modifies the spike raster diagram in panel
A to arrange the spike trains on the vertical axis according to the
$x$-coordinate of the position of the emitting neuron in physical
space. In contrast to the regular spike raster, we observe inhomogeneous
spatiotemporal features in network activity. The spatially binned
spike counts along the temporal axis in panel D, however, do not reveal
any unexpected structure. Thus, taking into account one coordinate
of the neurons in physical space hints at some organization of neuronal
activity. Nevertheless, a higher-dimensional analysis seems to be
required to uncover its origin, as the features of spatiotemporal
patterns can only be conjectured in 2D.

The emergence of planar wavelike spiking activity in 2D networks
was shown by \citet[Figures 3-5]{Voges2012}, but a 2D spatial visualization
of the data could not faithfully capture intermediate mixed patterns
such as rings and spiraling waves. Temporal snapshots of spatial activity
show the evolution of patterns, as seen in \citet[Figure 5]{Mehring2003},
\citet[Figures 2,12]{Yger2011}, \citet[Figure 6]{Voges2012} and
\citet[Figure 1]{Keane2015}. Series of such snapshots combined in
an animation or movie can be informative, but require settings to
be defined beforehand, leaving only little room for interactivity.
With such non-interactive visualization methods, crucial decisions
about a figure or an animation thus have to be made before a sufficient
intuition about the data exists. Flexible, interactive visualization
techniques identifying relevant dynamical features present in the
data would have the potential to avoid the tedious and time-consuming
loop of refining parameters and regenerating snapshots, animations
or movies. In addition, high-dimensional and multi-modal data demand
adequate workflows for analysis, from raw data to statistical measures,
where interactive visual analysis methods can play a major role. It
is for example essential to get a basic understanding of the datasets
to better decide what statistical methods to use for more elaborate
analysis. Furthermore, interactive visualization allows for explorative
data analysis, including dimensionality reduction of complex datasets,
highlighting of data points, and direct changes to visualization parameters.

For the development of supportive visual analytics tools, \citet{Shneiderman1996}
introduced the so-called \textit{\emph{\textquoteleft information-seeking
mantra}}\textquoteright . It describes the steps of common visual
analysis workflows: \textquotedblleft overview first, zoom and filter,
details on demand\textquotedblright . The first step provides a superficial
\textquoteleft overview\textquoteright{} of the data. In the second
step, \textquoteleft zooming\textquoteright{} into the dataset allows
the user to get a more detailed view on a chosen data subset. Application
of \textquoteleft filters\textquoteright{} implies a change in dimensionality
of the data or the extraction of particular features. Finally, \citet{Shneiderman1996}
proposes that visualization tools should enable the user to access
all details of selected data points.

To not restrict the user to only one visual representation of the
data, \citet{Wang2000} established the concept of \textquoteleft coordinated
multiple views\textquoteright . Coordinated multiple views is a paradigm
for the implementation of visual analysis applications that \textquotedblleft use
two or more distinct views to support the investigation of a single
conceptual entity\textquotedblright{} \citep[p. 110]{Wang2000}, and
has been applied in various contexts \citep[see for example][]{Roberts2007}.
Basic coordination of views addresses selection operations (e.g.,
whether to display only a subset of the data) and also includes immediate
control over animated frames (e.g., animation time step and playback
speed for time-resolved data). In addition, each view may have an
exclusive (view-specific) set of user controls and settings.

The activity exhibited by our example network is characterized by
a non-trivial interplay between neuronal populations resulting in
non-stationary activity in time and space. The neuroscientist needs
to identify the propagation of spiking activity within and across
individual layers over time and space, and simultaneously observe
population activity measures such as the LFP. This is an opportunity
to exercise the concepts by \citet{Shneiderman1996} and \citet{Wang2000}.
Visualization in most cases focuses on a specific aspect or hypothesis
to be tested by analyzing the corresponding data. Therefore, for each
task the neuroscientist combines a different set of views. Sometimes
particular views are not among the regularly used ones provided by
the visualization framework but are created ad hoc specifically for
the research question or the experimental protocol. Therefore the
analysis software environment needs to facilitate fast prototyping
of visualizations and an interface to a computing programming language
used in the scientific domain. This focus on a specific aspect under
investigation by the neuroscientist necessarily entails an individual
level of reduction or aggregation of the data. A particular visualization
realizes this preprocessing of the data with methods like binning
of data points in time or space, or by filtering out a certain subset
of parameters of each data point. For instance, the spike raster plot
in \prettyref{fig:raster}A displays the individual spikes of all
neurons whereas the bar chart in \prettyref{fig:raster}E visualizes
the total number of spikes per time step. The visualization abstracts
from the spikes of individual neurons and turns the focus to the whole
population. On the one hand the visualization simplifies interpretation
by presenting less detail, on the other hand the reduction increases
the chance of wrong or inaccurate conclusions. Historically, \citet[Fig. 4 middle]{Vaadia88}
illustrate a potential misinterpretation of the PSTH due to variability
in the onset of the neuronal response: a neuron observed over multiple
trials exhibits in the PSTH a smooth increase in spike rate, whereas
the raster plots show in each trial an abrupt increase in spike density
with a variable onset. \citet[Fig. 8]{Gruen02_81} demonstrate how
such misalignments can propagate to measures of statistical significance:
with respect to one trigger event the data show surplus spike synchrony
simply due to non-stationarity of spike rate, whereas with respect
to another trigger the rate is stationary and no excess synchrony
is detected. However, if a multi-view approach is implemented that
combines various visualizations, more than one aspect of the data
(more then one visual representation of differently processed data)
can be inspected simultaneously and can be put into relation. By interactive
addition and removal of certain views, this process can be made flexible
and thus address changes in the analysis goals or to consider findings
during the analysis. Finally, we require a solution that allows for
integration with platform-independent web-based technologies to keep
the accessibility of the tool as high as possible.

A variety of coordinated multi-view applications for the interactive
analysis of activity data has been described in literature, which
generally follow the information-seeking mantra. For models of neuronal
systems, the NEURON simulation environment \citep{Carnevale2006}
provides a graphical user interface based a modified version of the
discontinued InterViews library in addition to scripting in HOC and
Python \citep{HinesDavisonMuller2009}. The software itself offers
the possibility of drawing multiple concurrent windows with dynamic
and interactive plots of voltages, currents, morphology shapes and
phase planes that are updated while simulations of single-neuron models
or neuron networks are running. 3D visualization is not directly supported,
but NEURON's Python bindings also allows running simulations to interact
with modern visualization software, as for example incorporated by
NeuronVisio \citep{MattioniCohenLeNovere2012} that relies on the
OpenGL-accelerated Mayavi visualization toolset \citep{ramachandran2011mayavi}.
The simulation software for large-scale neuronal network models NEST\footnote{\href{http://nest-simulator.org}{http://nest-simulator.org}}
(NEural Simulation Tool, \citealp{Gewaltig_07_11204}) does not provide
built-in interactive visualization. The original authors state this
in their first report \citep{SYNOD} as a design decision based on
two considerations. First, in \citeyear{SYNOD} the life time of graphics
frameworks and libraries appeared much shorter than the envisioned
period of relevance of a simulation code. Thus, only a software stack
with a strict separation of levels would ensure platform independence
and sustainability of NEST. Second, a basic idea of the project is
to contribute to a software environment for \textquoteleft in virtu\textquoteright{}
now often called \textquoteleft in silico\textquoteright{} experiments
\citep[restated in][]{Diesmann01}. In this concept, the authors state,
simulated data and experimental data should be analyzed with the same
analysis tools to maximize comparability and reproducibility. At the
same time researchers at the department of Physiology and the Center
for Neural Computation of the Hebrew University in Jerusalem started
to work on an integrated analysis and visualization platform based
on Open Inventor\footnote{\href{https://www.openinventor.com}{https://www.openinventor.com}}
called Neural Data Analysis (NDA) but the project was abandoned with
the advent of MATLAB \citep{Vaadia_privcomm}. Recently \citeauthor{Nowke2013}
took on the challenge to develop a simulator independent visualization
platform for brain-scale neuronal networks. The VisNEST \citep{Nowke2013,Nowke2015}
framework visualizes the spiking activity of multi-area network models
\citep[using as an example][]{Schmidt2017} in a virtual environment.
The time-resolved activity data is mapped onto a 3D brain model. This
enables the researcher to interact with the model in 3D to expose
otherwise occluded parts of the brain and to relate brain activity
to anatomy. In a different view, a dynamic 3D graph represents time
course of spike exchange between different cortical areas. These representation
of spatial information can be combined with classic charts such as
spike raster plots. The tool does presently not account for the spatial
organization of activity within brain areas. Apart from VisNEST, other
standalone interactive multi-view applications have been developed
for simulated spiking data, for instance SNN3DViewer \citep{Kasinski2009}
and ViSimpl \citep{Galindo2016}. SNN3DViewer focuses on 3D neuronal
networks by visualizing individual neurons and their connections schematically,
including interactive control over the 3D visualization (navigation,
scale). ViSimpl combines a 3D particle-system-based visualization
of the simulated neuronal network using color coding for the activity,
supplemented by a set of data charts for single neurons and populations.
Geppetto\footnote{\href{http://www.geppetto.org}{http://www.geppetto.org}}
is a web-based modular platform for visualization and simulation of
complex biological systems including spiking neuronal networks. Unlike
the visualization concepts along which these tools have been developed,
we here focus on concepts that expose the spatial organization of
neuronal activity in layered networks and scale to signals from several
square millimeters of brain surface.

Beside the aforementioned softwares applicable with spike data, general-purpose
multi-view frameworks exist with different design goals and contexts
of use \citep{Roberts2007}. One generic high-level example is GLUE\footnote{\href{http://glueviz.org}{http://glueviz.org}},
a Python and OpenGL-based multi-view framework. Another powerful framework
is the now neglected OpenDX\footnote{\href{http://www.opendx.org}{http://www.opendx.org}}.
In contrast to the GLUE toolset, we here aim at web-based visualization.
Easy access to libraries of common plotting functions and methods
(scatter, line, surface plots etc.) is provided for most common programming
languages (C++, Python, MATLAB, etc.). Nevertheless, a large amount
of time and resources is still required to construct fully interactive
visualization tools adhering to the principles outlined by \citet{Shneiderman1996}
and \citet{Wang2000}. Including interactivity and time synchronization
between different visualizations may be demanding in terms of software
design and development time, however, existing plotting libraries
can be used to realize the individual visualizations.

As a reference implementation of our conceptual study, we introduce
the interactive visualization tool VIOLA, an open-source, platform-independent
and lightweight web-browser application. The tool is designed for
initial visual inspection of massively parallel data generated primarily
by simulations of spiking neuronal networks similar to the example
network illustrated in \prettyref{fig:network_sketch}. VIOLA is designed
around the information-seeking mantra and the concept of coordinated
multiple views. 2D and 3D visualizations support the exploration of
neuronal activity across space and time. The software can display
raw spiking output as well as spatiotemporally binned data that may
represent instantaneous spike counts gathered from nearby groups of
neurons. Spike and LFP data can be displayed simultaneously, thus
allowing for a multi-modal analysis. 

The manuscript is organized as follows: In \nameref{sec:results}
we present different visualization types and their application. Subsequently
(\nameref{sec:methods}) we describe their implementation in the visualization
tool VIOLA, the example network model and the phenomenological model
for the LFP signal. Finally in \nameref{sec:discussion} we conclude
our work and discuss general limitations of frameworks for explorative
visualization and potential future developments.

\section{Results}

\label{sec:results}

For the analysis of data, static figures can help to highlight certain
characteristics of the data or show results relevant for a particular
hypothesis. However, static figures hamper an exploratory analysis
of data as the adaption of data filters, visualization parameters,
or changes in the perspective (in case of 3D visualization) require
a re-rendering of the figure resulting in a very slow visual analysis
process. Interactive visualization tools tackle these shortcomings
by offering multiple views on the same data simultaneously, for example
by projecting the data across different dimensions. This allows the
user to investigate data at different levels of detail, and to adapt
visualization parameters in a dynamic and explorative manner as the
rendering of the visualization is continuously updated. Throughout
this section, we use the spike output of the point-neuron network
introduced in the \nameref{sec:introduction} as an example to demonstrate
appropriate visualization types in an interactive and multi-view framework. 
Neurons in this
network are placed in 2D sheets, and connections are drawn using distance-dependent
probabilities between pairs of neurons. The model represents spatially
heterogeneous neuronal activity across a $\unit[4\times4]{mm^{2}}$
cortical sheet. As we here focus on visualization methods in VIOLA,
we refer the reader to Sections \ref{sec:network_description}-\ref{sec:LFP}
for the details on our network implementation in NEST \citep{Kunkel2017},
Python-based preprocessing steps and predictions of a mesoscopic population
signal, the local field potential (LFP). We next describe in detail
the different views of VIOLA and their use cases.

\subsection{Views of VIOLA}

\label{sec:views}

VIOLA incorporates two conceptually different visualization types
with two separate \textquoteleft views\textquoteright{} each. The
first visualization type (view 1 and view 2 in \prettyref{fig:inst_data})
focuses on instantaneous snapshots of data across space. The second
visualization type (view 3 and view 4 in \prettyref{fig:time_series_data})
shows time series of data. We first present the visualizations of
preprocessed data described in \prettyref{sec:dataformats}. Views
1-3 may also be used to visualize raw data (non-preprocessed) as shown
in \prettyref{fig:raw}. 

\subsubsection{View 1: 2D spike-count rate}

\label{sec:view1}

\begin{figure}[t]
\includegraphics[width=1\textwidth]{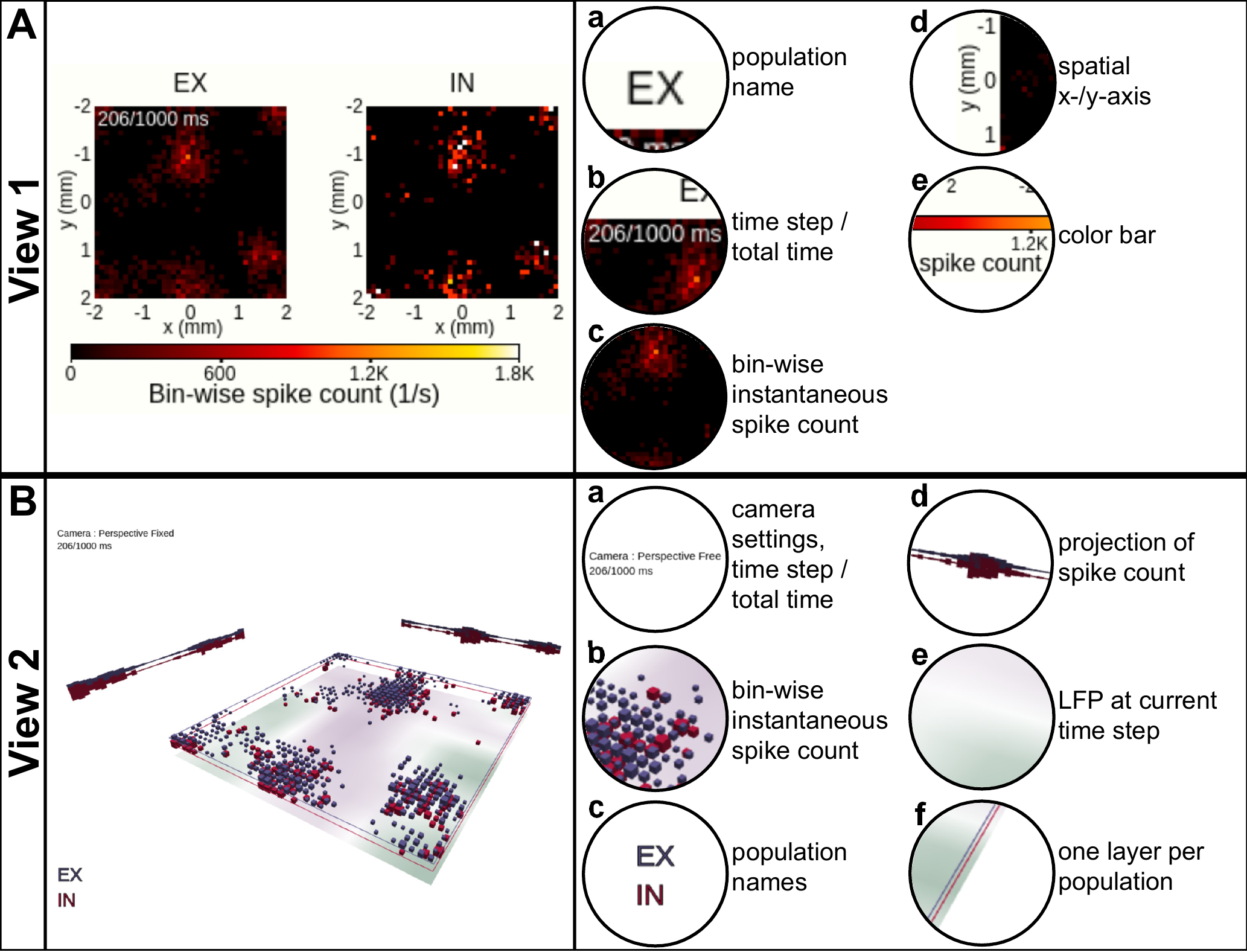} \caption{\textbf{View 1: 2D spike-count rate} Panel~A shows the instantaneous
spike-count rates, defined as the number of spikes per second occurring
within a spatiotemporal bin, using 2D image plots spanning the spatial
$x-$ and $y-$axes of the network layers. One separate image plot
is created for each network population and denoted by the population
name ($\mathrm{EX}$, $\mathrm{IN}$). The color map and corresponding
color bar for instantaneous spike-count rate values are shared among
all populations. In this and subsequent panels, we show the screen
shot of the view itself to the left and highlight its components to
the right. \textbf{View 2: 3D layered spike-count rate.} Panel~B
combines the data shown in Panel~A in a single 3D scene by stacking
the different population data on top of each other. The size of each
cubic marker denotes the magnitude of the corresponding bin-wise instantaneous
spike count, and its position corresponds to the spatial positions
of the respective bin. Unique colors are assigned to each layer as
indicated by the population names. Projections of the spike counts
along the $x-$ and $y-$axes are displayed towards the corresponding
edges. The optional bottom image plot layer shows the spatial variation
in an LFP-like signal at the present time step of the rendering loop. }
\label{fig:inst_data} 
\end{figure}

The 2D spike-count rate view (\prettyref{fig:inst_data}A) 
shows instantaneous activity data in
separate sub-panels for individual populations. The values in each
panel correspond to the instantaneous spike-count rate $\nu_{\beta}$
in one discrete spatiotemporal bin indexed by $\beta=(l_{x},l_{y},k)$
in our preprocessed data format ($l_{x}$ and $l_{y}$ denote spatial
bin indices along the $x-$and $y-$axes, and $k$ denotes a temporal
bin index). In this format, each spike event is added to the corresponding
spatiotemporal bin as described in detail in \prettyref{sec:dataformats}.
The color bar denotes bin values in units of spike counts per second
($\mathrm{spikes/s}$) and is shared between all sub-panels. This
view provides a side-by-side comparison of the spatially resolved
activity in each individual population. For a larger number of populations
(than shown here), it is, however, difficult to relate the activity
of one population to another population by visual observation. This
problem can however be amended by combining multiple population activities
in a single scene.

\subsubsection{View 2: 3D layered spike-count rate}

\label{sec:view2}

In the 3D layered spike-count rate view (\prettyref{fig:inst_data}B), 
we combine the activity
of all network layers in one 3D-animated scene. The view incorporates
the possibility to show also other activity measures, for example
the population LFP. The layers in view 2 correspond to the different
sub-panels for each population in view 1. Different populations are
here assigned unique colors. We chose to illustrate instantaneous
spike-count rate $\nu_{\beta}$ by dynamically sized cubic boxes.
The box sizes are by default scaled such that their volumes are proportional
to $\nu_{\beta}$ at each time step, thus low-activity bins may still
be visualized simultaneously with high-activity bins.

View 2 offers multiple possibilities for interactive adaptation of
the visualization. As suggested by the information-seeking mantra
\citep{Shneiderman1996}, the user can manually select which part
of the data to show, for example by switching on or off individual
layers that may occlude visibility of activity in other layers or
by setting the horizontal $x-$ and $y-$limits of the layers. It
is also possible to reduce the opacity of the colored boxes or to
scale their side length linearly. The camera can be set either to
an orthographic or perspective-corrected projection mode. Dependent
on the projection mode, the camera can be moved freely and allows
for zooming, panning and rotating the scene. One can easily reset
the camera to its default position by the click of a button or select
different preset camera positions such as on top or to the side.

The major benefit provided by view 2 over view 1 is the possibility
to visually relate the activity in one layer to other layers as all
layers are drawn in the same 3D scene. As box volumes are computed
from instantaneous spike-count rate values, this view brings the attention
of the user to spatial regions of the network with high local activity.
While views 1 and 2 offer flexible visualizations of instantaneous
activity across space, we next consider scenes capable of showing
time-series data.

\subsubsection{View 3: Scrolling spike-count rate plot}

\label{sec:view3}

\begin{figure}[t]
\includegraphics[width=1\textwidth]{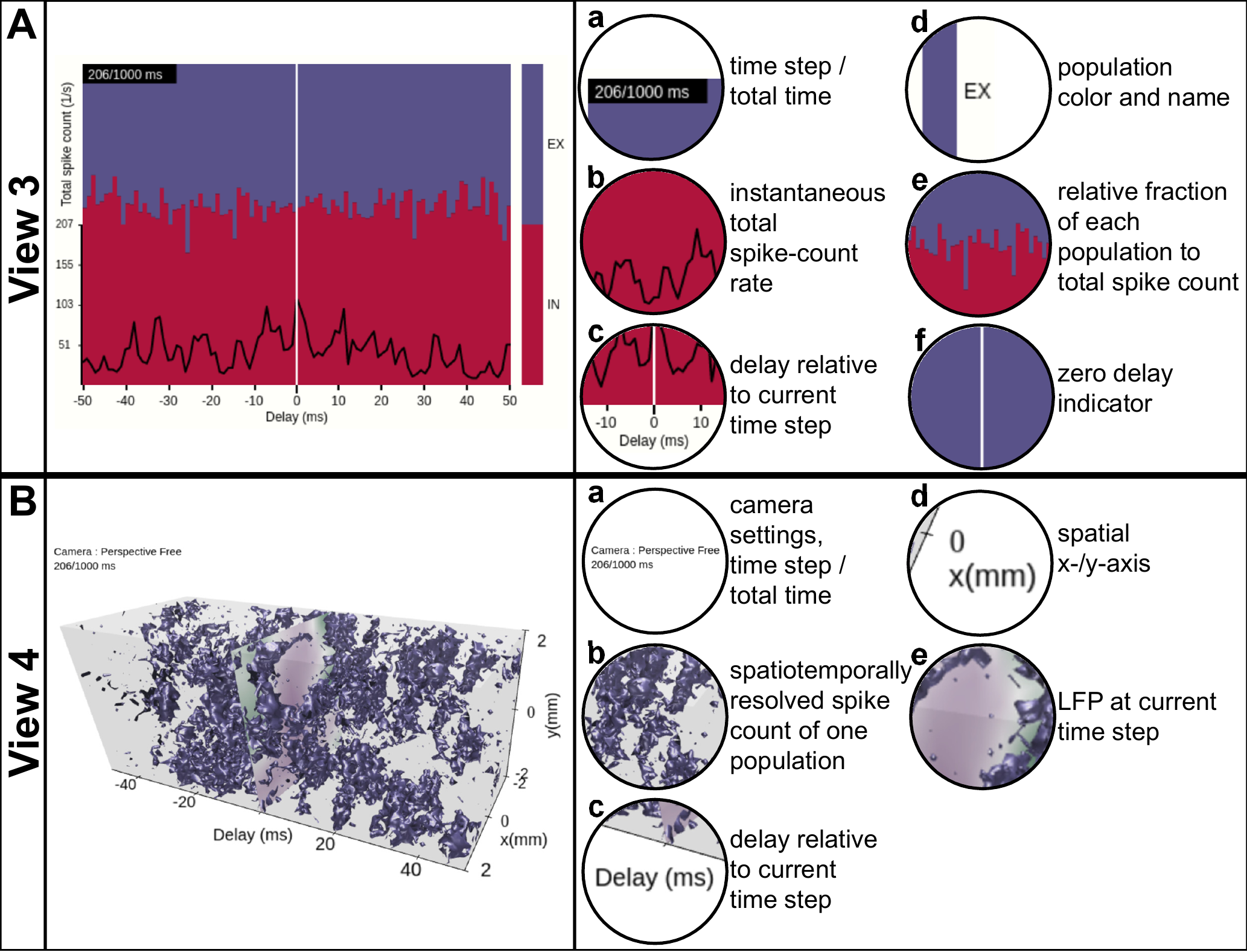} \caption{\textbf{View 3: Scrolling spike-count rate plot.} Panel~A is a time-series
representation of the data across a user-variable time interval around
the present time step of VIOLA's rendering loop, indicated by the
vertical white line. The instantaneous total spike-count rate summed
over all populations is drawn using a black line. The relative fraction
of spikes of each population to the total spike count is shown as
a stacked, normalized histogram. The population outputs are color
coded as in view 2. \textbf{View 4: Scrolling spike-count rate iso-surface
plot.} Panel~B provides a 3D representation of the spatiotemporally
resolved spike-count rates of one selected network population across
a user-variable time window. The spike-count rate is rendered as a
closed iso-value surface in the color of the respective population
and extends in both space ($x-$ and $y-$axes) and time (delay axis).
The present time step in the visualization is indicated by a time
lag of zero on the time-delay axis. At zero time delay we also show
the LFP signal corresponding to the present time step in the animation. }
\label{fig:time_series_data} 
\end{figure}

The scrolling spike-count rate plot (\prettyref{fig:time_series_data}A) 
is a time-series representation
of the data that neglects spatial features of the network and its
activity. It shows the time evolution of the total spike-count rate
$\nu_{k}$ (black line), defined as the sum over the spike-count rates
of all spatial bins and populations divided by the number of bins,
together with the relative rate of each individual population (colored
stacked plot). $\nu_{k}$ is defined as $\nu_{k}\equiv1/(L_{x}L_{y})\cdot\sum_{X}\sum_{l_{x}}\sum_{l_{y}}\nu_{\beta}$
with $\beta=(l_{x},l_{y},k)$ and where $L_{x}$ and $L_{y}$ denote
the number of bins along the $x-$ and $y-$axes and neuronal populations
are denoted $X$. The per-population spike-count rates can therefore
be inferred by multiplication of the total rate with the fraction
of spiking in individual sub-populations. The color coding of each
population corresponds to the one used in view 2, but can also be
read from the bar to the right. The plot is centered on the current
time step (vertical white line indicator) when scrolling through data
points in the animation. It allows for interactive change of the width
of the visible time window and also permits to manually select or
deselect individual populations to be displayed. View 3 provides a
temporal overview of the data and allows to identify time intervals
of interest, for example, due to an external perturbation.

\subsubsection{View 4: Scrolling spike-count rate iso-surface plot}

\label{sec:view4}

The instantaneous spike-count rates of our example network are time-series
activity data with 2D spatial structure. In order to visualize such
data without loss of dimensionality, a 3D representation is in general
required (unlike for instance view 3). The scrolling spike-count rate
iso-surface scene in \prettyref{fig:time_series_data}B 
simultaneously shows the evolution
of network activity in space (as in views 1 and 2) and time (as in
view 3). The rate iso-value surfaces of each population is rendered
using the computer-graphics algorithm \textquoteleft marching cubes\textquoteright{}
\citep{Lorensen1987}. The color coding of the individual populations
matches the coding used in views 2 and 3. In terms of user interactivity,
the user can set the threshold (isolation) for the surfaces. Furthermore,
the user can select which populations to show, vary their opacity
level, apply a temporal offset to individual populations, change the
width of the time window, and take full control of the viewpoint in
the 3D scene as in view 2.

\subsubsection{Raw data views}

\label{sec:raw}

\begin{figure}[t]
\includegraphics[width=1\textwidth]{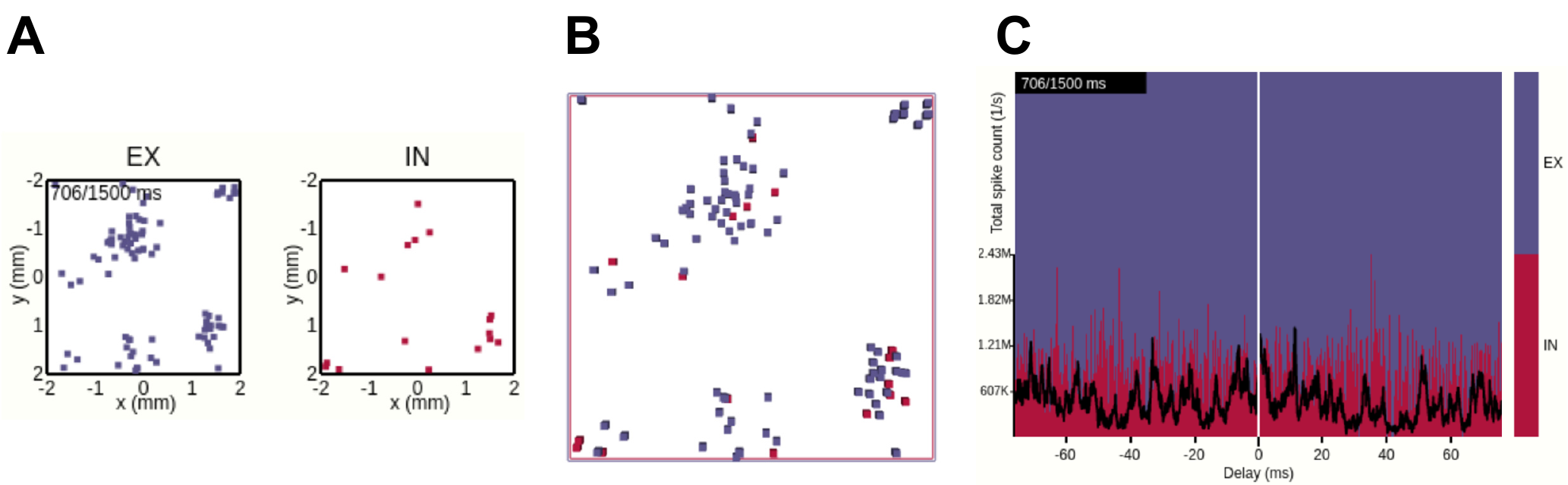} \caption{\textbf{Views 1-3 applied with raw data formats.} \textbf{A}~View
1: Each dot corresponds to a single spike event of a neuronal unit
at its spatial location in the network. \textbf{B}~View 2: Perspective
top-down view onto stacked layers of Panel~A. \textbf{C}~View 3:
The stacked plot has the temporal bin-size of the simulation resolution.
The black trace shows the spike count of all neurons per time bin
(in units of $\mathrm{spikes/s}$). }
\label{fig:raw} 
\end{figure}

In addition to visualizing spatiotemporally binned, preprocessed data,
views 1-3 can also be used with raw simulation output files formatted
according to the description in \prettyref{sec:dataformats}. With
raw file output, view 1 (\prettyref{sec:view1}, \prettyref{fig:raw}A)
displays for each animation time step a square marker for each spike
time $t_{j}^{s}$ at the spatial location $(x_{j},y_{j})$ of neuron
$j$ in population $X$. The square marker color is population specific.
Likewise, view 2 (\prettyref{sec:view2}, \prettyref{fig:raw}B) shows
boxes of equal size for each spike event, colored according to population.
The view allows, as with precomputed spike-count rates, to show spiking
activity in each population in the same scene. We here show a snapshot
of the spiking activity in perspective mode, and top-down. The main
interactive feature of views 1 and 2 incorporated with raw data files
is the option to reduce the neuron density to be displayed. As with
preprocessed data visualization with view 2, individual layers can
be activated/deactivated, one can switch between the orthographic
and perspective viewing modes, and the camera can be positioned freely.
View 3 (\prettyref{sec:view3}) applied to raw data is shown in \prettyref{fig:raw}C.
The temporal bin size of the animation is then equal to the simulation
time step $dt$ (one spike therefore results in the spike count rate
$1/dt$ in units of $\mathrm{spikes}/s$ for that instant). The total
spike count (black line) is summed over all neurons, in contrast to
the mean over per-bin rates as in the case of the preprocessed data.
The relative spike count per population is shown as a stacked plot
normalized by the total amount of spikes in each temporal bin.

\subsection{VIOLA use case}

\label{results:usecase}

Numerical model development representing a physical system comprises
implementation, simulation, analysis as well as comparison, validation
and verification steps. Such model development is important for building
hypotheses and aiding interpretation based on experimental data and
observations. We here demonstrate how the views described above can
be integrated with the development of a spiking point-neuron network
model. For this purpose, we can hypothesize that transient external
input to a layered spiking point-neuron network model with distance-dependent
recurrent connections results in propagating spatiotemporal activity.
We wish to assess the spatial extent and temporal duration of the
network response to external perturbation and whether or not the unperturbed
network state is recovered. In this use case we do this assessment
by visual inspection prior to any detailed numerical analysis, focusing
on the importance of coordinated multiple views \citep{Wang2000,Roberts2007}
and Shneiderman's information-seeking mantra \citep{Shneiderman1996}.
For our hypothesis above we will therefore use our implementations
of views 1-4 in VIOLA to rapidly analyze our network activity. We
show that a combination of the different views is needed to asses
the relevant aspects in the data, which is the evoked response to
a network perturbation.

The layered point-neuron network illustrated in \prettyref{fig:network_sketch}B
consists of an excitatory ($\mathrm{EX}$) and an inhibitory ($\mathrm{IN}$)
neuronal population plus one stimulus population ($\mathrm{STIM}$)
. Each neuron is placed randomly within square sheets. $\mathrm{EX}$,
$\mathrm{IN}$ and $\mathrm{STIM}$ units are connected using distance-dependent
rules as illustrated in \prettyref{fig:network_sketch}A. The connectivity
is periodic across boundaries (torus connectivity). The detailed network
description is given in \prettyref{sec:network_description}. The
main simulation output is spike times of individual neurons, neuron
locations and a synthetic LFP signal (see \prettyref{sec:LFP} for
details). Our initial preprocessing steps and corresponding data formats
are described in \prettyref{sec:dataformats}.

\subsubsection{Temporal features of evoked network activity}

\label{results:part2}

\begin{figure}[t]
\includegraphics[width=1\textwidth]{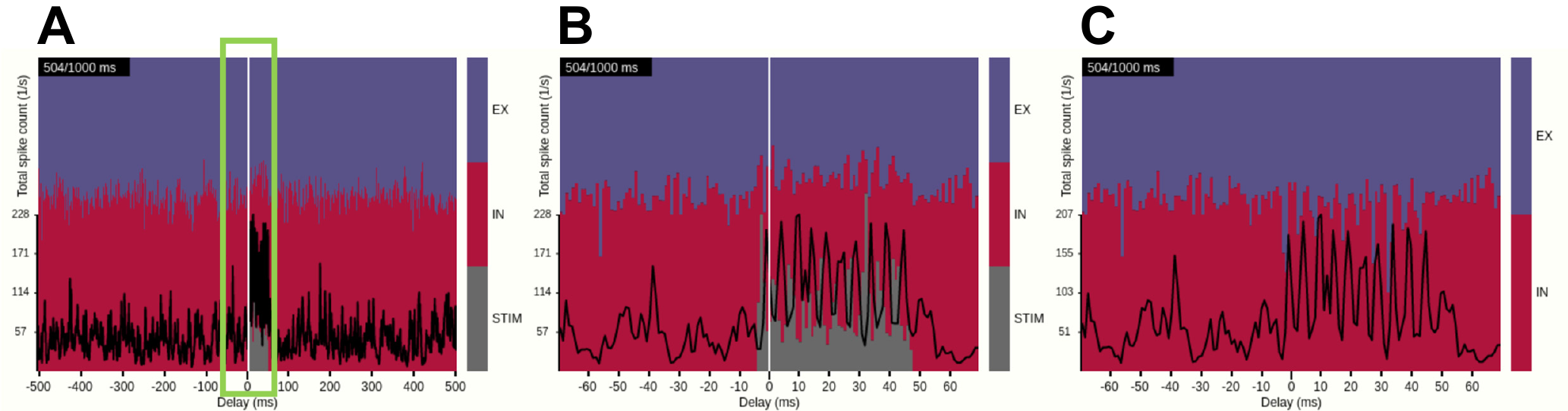} \caption{\textbf{Identifying a time interval of interest with view 3.} \textbf{A}~The
spike-count rate summed across spatial bins and the relative contributions
by populations $\mathrm{EX}$, $\mathrm{IN}$ and $\mathrm{STIM}$
shown for a time window of $\unit[\pm500]{ms}$ around the current
time step of the animation. \textbf{B}~Same as Panel A, but with
a narrower time window of $\unit[\pm70]{ms}$ (indicated by the green
frame in panel A), highlighting the activation of the $\mathrm{STIM}$
population and corresponding network response. \textbf{C}~Same as
panel B, but with the $\mathrm{STIM}$ contribution turned off. }
\label{fig:use_case_graph} 
\end{figure}

We first focus on ongoing activity of the network in the time domain,
as provided by view 3. This view implements a scrolling spike-count
rate plot which ignores spatial information. Interactive control of
the view's time window allows for quick identification of events of
interest from the full duration of the simulation (\prettyref{fig:use_case_graph}A).
One such event that is clearly differentiated from other ongoing activity
is the activation of the external $\mathrm{STIM}$ population at the
animation time step of $\unit[500]{ms}$. Pausing the animation at
$\unit[504]{ms}$ and zooming in onto the event (\prettyref{fig:use_case_graph}B)
allows for a detailed look on how the total spike-count rate (black
trace) increases and oscillates while the stimulus is active, and
confirms that the stimulus duration was $\unit[50]{ms}$. The color-coded
stacked histogram reveals that during stimulus activation a large
relative fraction of spike events is contributed by the $\mathrm{STIM}$
units (gray), while the relative fraction generated by the recurrently
connected $\mathrm{EX}$ (blue) and $\mathrm{IN}$ (red) units is
reduced. We may also conclude that the transient onset of the stimulus
results in temporally brief imbalances between excitatory and inhibitory
populations in the network as the relative rate of the inhibitory
population drops with regular intervals during the stimulation period.
The imbalances occur at the stimulus onset and during each period
of the resulting network oscillation (from recurrent interactions
between excitatory and inhibitory neurons). This network spike-rate
imbalance is even more pronounced when the $\mathrm{STIM}$ activity
is hidden (\prettyref{fig:use_case_graph}C). We note, however, that
the rate balance averaged over the stimulus duration is similar to
time-averaged rate balance in the non-perturbed state.

From the visualization we can also infer that the external perturbation
to the network does not result in a shifted network state after the
stimulus is switched off. Overall rate fluctuations and relative fractions
of spike-count rates appear comparable before and after the stimulus
period, unlike networks that may display multi-stable patterns of
activity \citep{LitwinKumar2012,Miller2016} wherein their state can
shift from one attractor to another either spontaneously or due to
a perturbation.

\subsubsection{Spatial features of evoked network activity}

\label{results:part3}

\begin{figure}[t]
\includegraphics[width=1\textwidth]{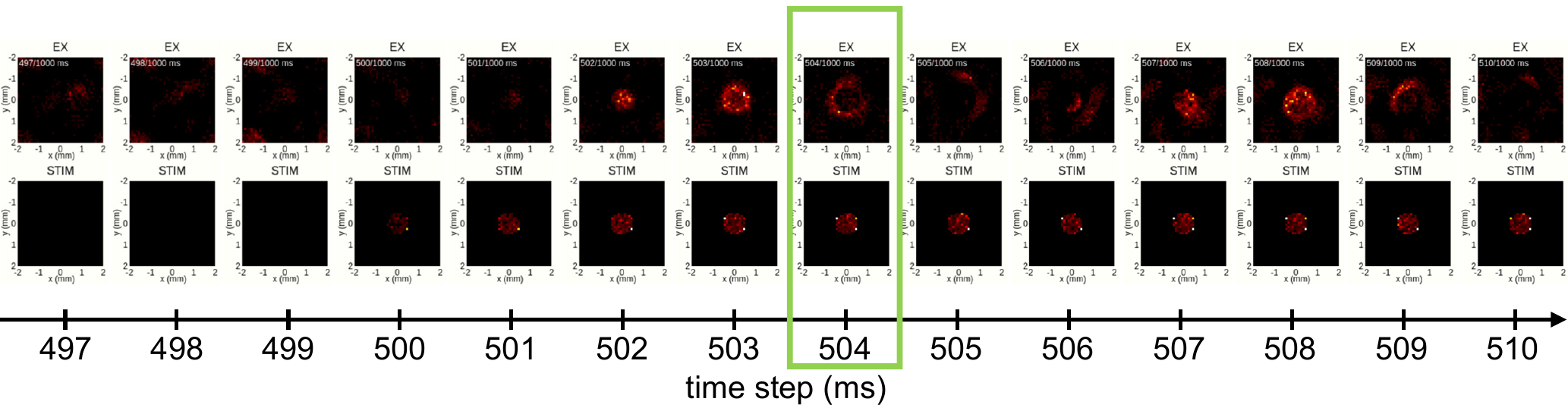} \caption{\textbf{From spontaneous to evoked activity, resolved in time and
2D space.} Time frames of populations $\mathrm{EX}$ (first row) and
$\mathrm{STIM}$ (second row) are captured from view 1 every $\unit[1]{ms}$.
After three frames showing spontaneous activity of population $\mathrm{EX}$,
the $\mathrm{STIM}$ layer is activated (first visible in the fourth
column, at $\unit[500]{ms}$), resulting in a repeated pattern of
outward spread of activity in the $\mathrm{EX}$ layer. The time step
highlighted by a green outline (at $\unit[504]{ms}$) corresponds
to the animation time step in \prettyref{fig:use_case_graph}. }
\label{fig:use_case_frames} 
\end{figure}

Having identified a time segment of particular interest (the stimulus
duration), we next exploit view 1, the 2D spike-count rate view, and
focus on spatial aspects of the evoked network activity. \prettyref{fig:use_case_frames}
shows a series of snapshots from the instantaneous spike-count rate
animation across space for the $\mathrm{EX}$ (top row) and the $\mathrm{STIM}$
(bottom row) layers. Snapshots are shown for successive bins of width
$\Delta t$. The first three columns in \prettyref{fig:use_case_frames}
show spontaneous activity of the $\mathrm{EX}$ units. Thereafter
the $\mathrm{STIM}$ population is switched on, as seen in the fourth
column of the bottom row. The activity of the STIM layer is by construction
confined to a circle at the center of the network. As the activity
of the $N_{\mathrm{STIM}}$ units in the $\mathrm{STIM}$ layer is
governed by Poisson processes with rate expectations $\nu_{\mathrm{STIM}}$,
its spike intensity remains fairly constant (except for the bin at
$\unit[500]{ms}$ as the time bin is centered on the time step). In
layer $\mathrm{EX}$, the stimulus elicits an increase in activity
spreading outwards from the center. This response dies out after a
few milliseconds due to recurrent inhibition, but reoccurs regularly
as reflected by the oscillatory behavior observed in \prettyref{fig:use_case_frames}B.
The time step at $\unit[504]{ms}$ highlighted by the green outline
is the same as in \prettyref{fig:use_case_frames} and latter Figures
\ref{fig:use_case_layers} and \ref{fig:use_case_timeline}.

\subsubsection{2D and 3D views of spatial activity}

\label{results:part4}

\begin{figure}[t]
\includegraphics[width=1\textwidth]{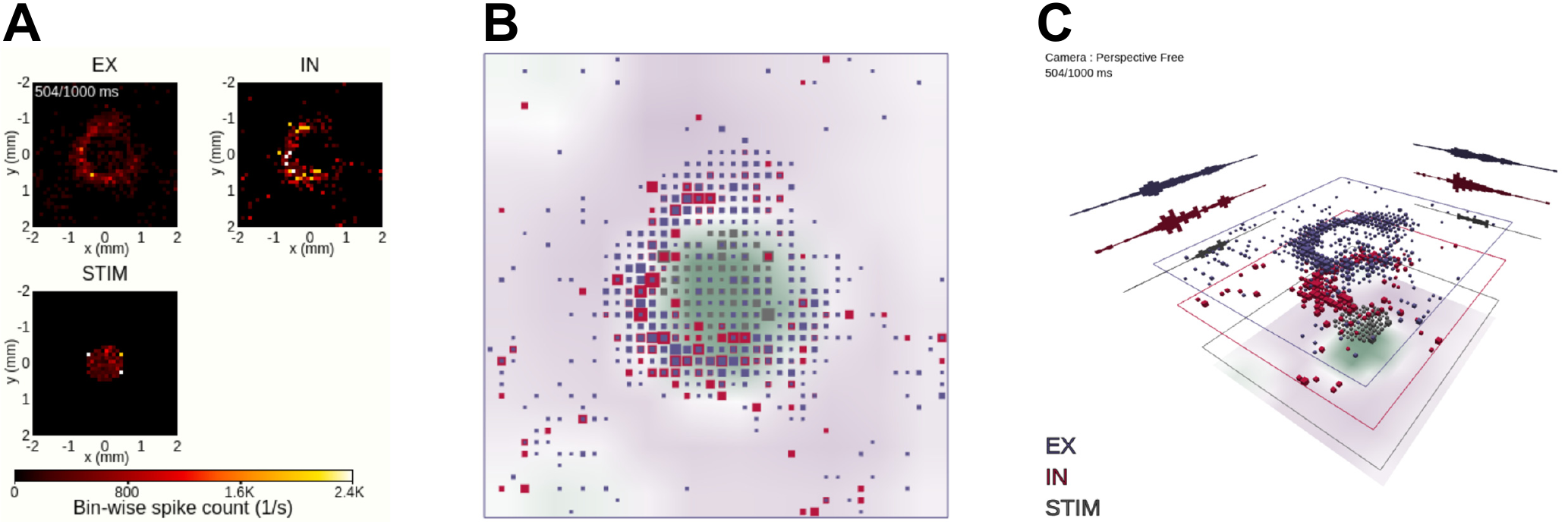} \caption{\textbf{Coordinated views on a temporal snapshot of the neuronal activity.}
\textbf{A}~Instantaneous spike-count rates in layers $\mathrm{EX}$,
$\mathrm{IN}$ and $\mathrm{STIM}$ using view 1. The animation time
step of $\unit[504]{ms}$ is identical to the one in Figs \ref{fig:use_case_graph}-\ref{fig:use_case_frames}
in this and subsequent panels. \textbf{B}~Orthographic top-down view
onto stacked population layers and LFP image plot with view 2. \textbf{C}~Perspective
view with large layer separation, including summed spike counts projected
towards the layer edges in view 2. }
\label{fig:use_case_layers} 
\end{figure}

In order to relate the spatial relationship between activity in individual
populations, we compare in \prettyref{fig:use_case_layers} three
different layer-wise animations of neuronal activity. View 1 (\prettyref{fig:use_case_layers}A)
shows individual 2D image plots for the spike-count rates per population,
with a shared color bar coding for instantaneous spike-count rate
values. This view offers an accurate spatial representation of network
activity in temporal bins of width $\Delta t$, showcasing the locality
of the $\mathrm{STIM}$ layer activity and the wider spread of evoked
activity in the $\mathrm{EX}$ and $\mathrm{IN}$ layers. This view
does not, however, offer interactive features except time control
of the animation (shared with views 2-4) and global scaling of the
color-value mapping (sensitivity control, shared with views 2 and
3).

The 3D-scene provided by view 2 adds additional interactive features
and incorporates the layer-resolved data of view 1 in one animation
(\prettyref{fig:use_case_layers}B,C). Panels B and C show the same
temporal snapshot of activity as in Panel A. The view shows also the
spatial variation of the LFP signal that we synthesized from network
activity. The LFP signal, here shown as image plot with a color-coding
reflecting its magnitude and sign, is more difficult to relate to
the ongoing activity, as it is inherently a signal driven by past
spiking activity (resulting of delayed synaptic activation on postsynaptic
neurons from spiking activity in presynaptic neurons, cf. \prettyref{sec:LFP}).
We then compare rate values of one spatial bin and one population
to other spatial locations and other populations through their different
color codings and cube sizes. An observation is that activity in the
$\mathrm{EX}$ and $\mathrm{IN}$ layers are typically confined within
the same spatial region of the network, while a larger fraction of
the network is quiescent at the time. This observation can for example
explain high variability in interspike-intervals of individual neurons
\citep{Keane2015}, as neurons may fire frequently while fronts of
activity spread across the network and remain quiet until the next
burst of activity.

In terms of using interactive features offered in view 2, we turn
off the orthographic mode of panel B and go back to its default 3D
perspective in panel C. We also rotate the viewpoint in order to directly
focus on highly active parts of the network. Furthermore, the different
layers of the network and LFP are offset vertically, with dynamic
projections of the sum of spiking activity across each respective
spatial axis for each network layer. From this setup of view 2, we
can better infer the activity in each individual layer, including
that of the LFP layer, without switching off individual layers.

\subsubsection{Spatiotemporally resolved network activity}

\label{results:part5}

\begin{figure}[t]
\includegraphics[width=1\textwidth]{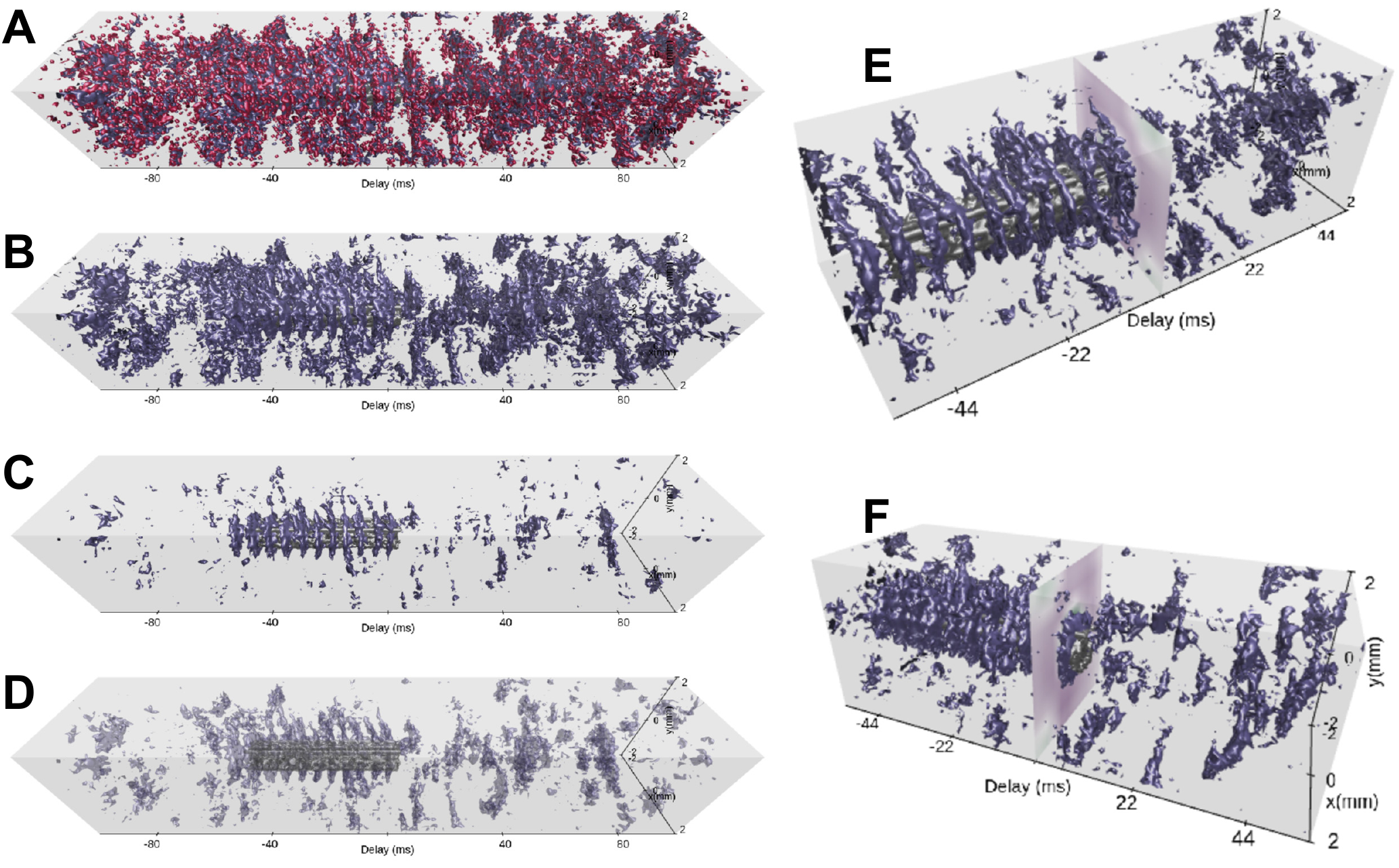} \caption{\textbf{Spatiotemporally resolved activity.} \textbf{A}~Spike-count
rates across time and space of populations $\mathrm{EX}$ (blue),
$\mathrm{IN}$ (red) and $\mathrm{STIM}$ (gray) shown with view 4
for a time window of $\unit[\pm100]{ms}$ around the present time
step of the animation. The isolation threshold is set to a rate of
$\unit[100]{spikes/s}$. The animation time step is identical to the
one in Figures \ref{fig:use_case_graph}-\ref{fig:use_case_layers}
in this and subsequent panels. \textbf{B}~Same as panel A, but with
$\mathrm{IN}$ activity turned off. \textbf{C}~Same as Panel B, but
with an increased isolation threshold of $360$ spikes per second.
\textbf{D}~Same as panel C, but with reduced opacity of layer $\mathrm{EX}$
activity and an isolation threshold of $\unit[195]{spikes/s}$. \textbf{E}~A
narrower time window ($\unit[\pm55]{ms}$) and shifted camera position
(isolation threshold of $\unit[195]{spikes/s}$). The image plot at
a delay of $\unit[0]{ms}$ shows the synthesized LFP signal across
space. \textbf{F}~Same as panel E, but with the camera position rotated
around the vertical $z-$axis. }
\label{fig:use_case_timeline} 
\end{figure}

We finally investigate network activity in space and time using the
3D scene provided by view 4. Similar to the scrolling spike-count
rate plot of view 3, view 4 allows full control of the time axis.
The activity of all populations $\mathrm{EX}$, $\mathrm{IN}$ and
$\mathrm{STIM}$ is displayed for a wide ($\unit[200]{ms}$) temporal
segment using red, blue and gray iso-surfaces, respectively, in \prettyref{fig:use_case_timeline}A.
We have centered the current time step (at $\unit[504]{ms}$) on the
evoked activity in the STIM layer (highlighted in \prettyref{fig:use_case_frames}).
It is already possible to identify activity patterns confined in space
and time. However, it remains difficult to assess how spontaneous
network activity changes in response to the stimulus due to occlusion
of one surface by another, an inherent issue with multiple solid surfaces.
In \prettyref{fig:use_case_timeline}B we therefore hide the activity
of the $\mathrm{IN}$ layer and focus on the activity in the $\mathrm{EX}$
layer. The surfaces correspond to the bin-wise instantaneous spike-count
rates at an isolation threshold of $\unit[100]{\mathrm{spikes/s}}$.
Increasing this threshold to $\unit[360]{spikes/s}$ (\prettyref{fig:use_case_timeline}C)
reveals that regular bursts of high rates occur at the center of the
layer, in the period when the $\mathrm{STIM}$ layer is activated.
In the other views, these bursts may be seen as rate oscillations
(\prettyref{fig:use_case_graph}) or pulsating spatial activity (Figures
\ref{fig:use_case_frames}-\ref{fig:use_case_layers}). We here show
that the attenuation of activity radiating outward from the center
is rather strong.

Using view 4, both the oscillation frequency and the outward spread
of activity in the $\mathrm{EX}$ population can be assessed. We highlight
the $\mathrm{STIM}$ activity by reducing the opacity of the $\mathrm{EX}$
surfaces in \prettyref{fig:use_case_timeline}D. This reduces occlusion
problems present with multiple overlapping opaque surfaces, and thus
allows relating the activity in these two populations to one another.
A smaller time segment of the scene is shown in panels E and F where
we also demonstrate different camera positions. Rotating the camera
allows us to observe the synthesized LFP signal at the current time
step, and the corresponding network interactions resulting in a strong
LFP fluctuation. We also observe the temporal offset between stimulus
onset and a response in the $\mathrm{EX}$ activity as shown in \prettyref{fig:use_case_timeline}F.

In contrast to the previously discussed applications of views 1-3,
the 3D-scene of view 4 allows to relate both temporal and spatial
aspects of the spiking activity of different neuronal populations
and the LFP signal to one another. With this view, we can get an overview
of a large time segment and several populations and then use its incorporated
interactive features in order to explore the network activity under
influence of the stimulus. The focus of this view lies on highlighting
qualitatively interesting features of the data on spatiotemporal scales
such as the oscillating activity of $\mathrm{EX}$ population surrounding
the $\mathrm{STIM}$ location (as in \prettyref{fig:use_case_timeline}C)
or the temporal offset between $\mathrm{STIM}$ and $\mathrm{EX}$
seen in \prettyref{fig:use_case_timeline}F. Views 1 and 3, however,
better resolve quantitative rate values or temporal offsets, respectively,
than views 2 and 4. 

\section{Methods}

\label{sec:methods}

\subsection{Data formats}

\label{sec:dataformats}

The data we consider for visualization are sequences $S_{j}$ of spike
times $t_{j}=\sum_{s\in S_{j}}\delta(t_{j}^{s})$ of a neuronal unit
$j\in X$ located at coordinate ($x_{j},y_{j}$), where $X$ denotes
a neuronal population of size $N_{X}$. Individual spike times $t_{j}^{s}$
are constrained to a discrete grid $n\cdot dt$ for $n\in\{0,1,2,...,n_{\mathrm{steps}}-1\}$,
where $dt$ is the time-resolution of spike acquisition and $n_{\mathrm{steps}}$
the number of time steps in the acquisition period $T$. We assume
that the raw spike data to be visualized is available in two pure
text files per population $X$. The first file contains two columns
with values separated by a white space. Its first column contains
integer numbers representing \textquoteleft global neuron identifiers\textquoteright{}
(neuron IDs) $j$, while the second column contains corresponding
spike times $t_{j}^{s}$ in units of $\mathrm{ms}$. This data format,
first introduced for experimental data and reviewed in \citet{Rostami2017},
is the default output format for spike data of the neuronal network
simulator NEST \citep{Kunkel2017}. While the floating point data
type is sufficient for displays and the computation of single-neuron
and population spike rates, the format is only safe for correlation
analysis if the time step is a power of two \citep[A.2]{Morrison2007}.
The latter guarantees that spike times have a representation in the
data type. An alternative is to use the original definition of the
format and denote spike times by the integers $n$, thus expressing
time in units of the resolution of the grid. The second file contains
three space-separated columns. Its first column contains unit IDs
$j$, while columns two and three contain the corresponding coordinates
$x_{j}$ and $y_{j}$ in units of $\mathrm{mm}$. NEST internally
represents networks as a graph where edges denote connections. Neurons
cannot be interrogated for their location and are only identified
by their ID, thus the information on the location must be defined
and stored explicitly.

We consider another text-based data format for the visualization of
spike data that are preprocessed by a temporal and spatial binning
procedure. For the temporal binning we define a temporal bin size
$\Delta t$ as an integer multiple of the acquisition time resolution
$dt$. For spatial binning of neuron positions along the $x-$ and
$y-$axes we define the bin widths $\Delta l$. The third spatial
dimension ($z-$axis) is ignored. Assuming an acquisition period $T$
and the side length $L$ of the centered square network domain, the
number of temporal bins is $K=T/\Delta t$ and the numbers of spatial
bins along each axis $\{L_{x},L_{y}\}=L/\Delta l$. A spatiotemporal
bin is indexed by the length-three tuple of indices $\beta=(l_{x}\in\{0,1,...,L_{x}-1\},l_{y}\in\{0,1,...,L_{y}-1\},k\in\{0,1,...,K-1\})$,
spanning $x\in[l_{x}\Delta l-L/2,(l_{x}+1)\Delta l-L/2)$, $y\in[l_{y}\Delta l-L/2,(l_{y}+1)\Delta l-L/2)$
and $t\in[k\Delta t,(k+1)\Delta t)$. In each spatiotemporal bin,
we sum for every population $X$ the number of spike events and divide
by the temporal bin size $\Delta t$. We refer to this measure as
the instantaneous spike-count rate $\nu_{\beta}$ in units of 1/s.
The preprocessed data is contained in one single file per population
with four space-separated columns. Indices $l_{x},l_{y}$ and $k$
for each spatiotemporal bin are put in columns 1, 2 and 3, respectively,
while the 4th column contains the corresponding rate value. Rows are
ordered in iteration running order according to $k\in[0,1,...,K-1]$
over all $l_{x}\in[0,1,...,L_{x}-1]$ and finally over all $l_{y}\in[0,1,...,L_{y}-1]$.
Row entries where $\nu_{\beta}=0$ is not written. The same data format
is used to represent the evolution of spatially organized analog data
with spatial resolution $\Delta l^{\phi}$. The unit of the data depends
on the actual measure, for example $\mathrm{mV}$ in case of the LFP.

\subsection{Reference implementation}

\label{sec:methods_implementation}

\begin{figure}[t]
\includegraphics[width=1\textwidth]{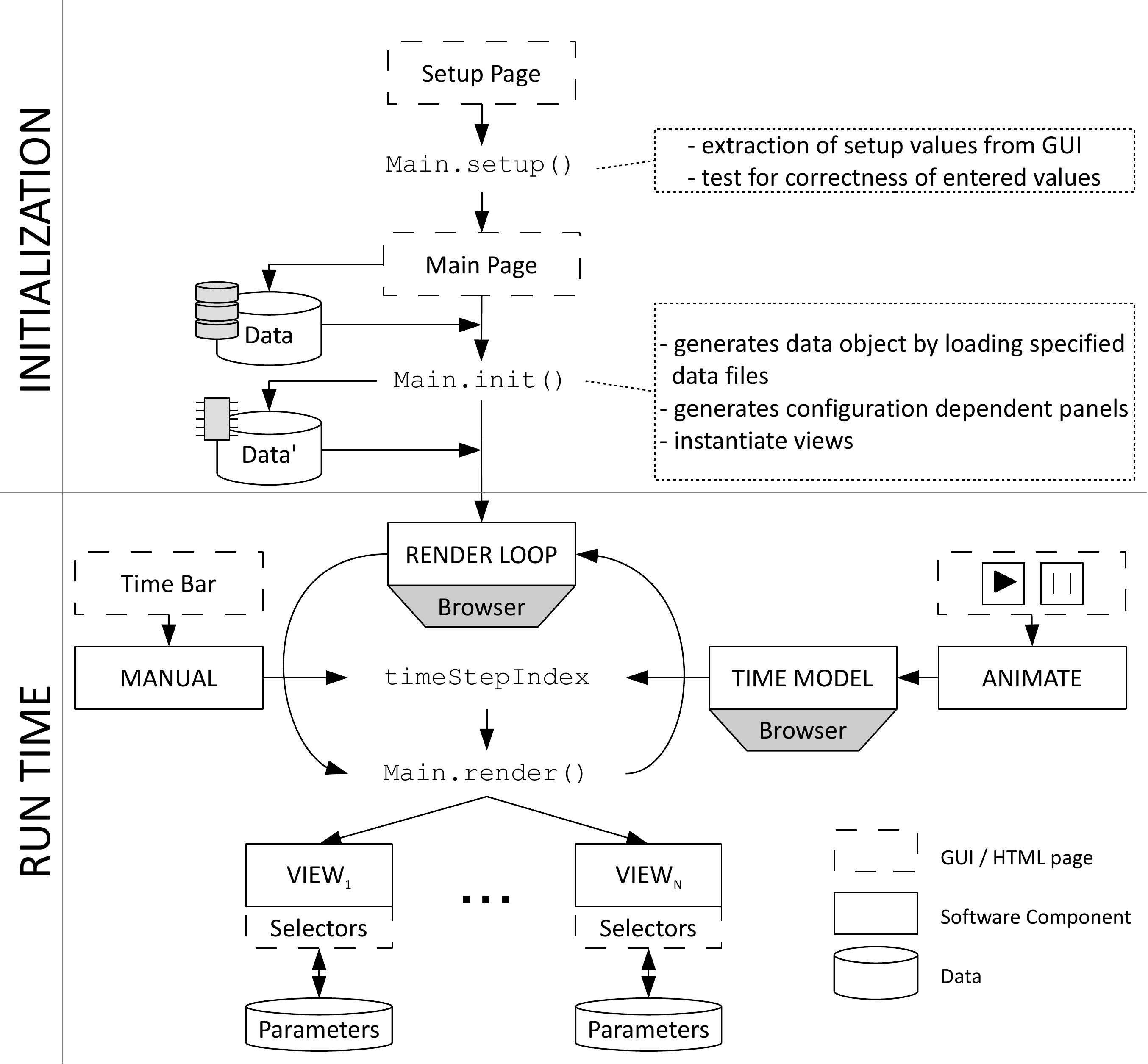}
\caption{\textbf{Flow chart of VIOLA's components.} VIOLA incorporates two
main parts: \textquoteleft initialization\textquoteright{} (top) and
\textquoteleft run time\textquoteright{} (bottom). The initialization
procedure defines a \texttt{Setup} and \texttt{Main Panel} in VIOLA's
GUI and the corresponding \texttt{Main.setup()} and \texttt{Main.init()}
functions. The \texttt{Main.setup()} function is used for setting
initial values, while the \texttt{Main.init()} function allows for
loading datasets with parameters that depend on setup values. The
\texttt{Main.init()} function also instantiates the different views
in the application. The run-time component of VIOLA uses a rendering
loop and time model provided by the web browser. The time model is
needed to synchronize the rendering of each view, such that at all
times each view shows the same time step of the data and thus compensates
for different redering times of the views. By default, the \texttt{timeStepIndex}
in the rendering loop is automatically updated by the browser (using
the time model), but can also be set by the user, e.g. by a slider
widget. The browser time model is controlled using the \textquoteleft animate\textquoteright{}
widget, while a time bar is used for \textquoteleft manual\textquoteright{}
time selection. Each update of the \texttt{timeStepIndex} triggers
an execution of the \texttt{Main.render()} method and a corresponding
update of all views. Parameters for the different views can be modified
during run time as each view offers its individual input widgets. }
\label{fig:violaflow} 
\end{figure}

We have made implementations of the visualization types discussed
throughout this paper available in the tool VIOLA (VIsualization Of
Layer Activity). \prettyref{fig:violaflow} illustrates the web-based
JavaScript framework integrating the different visualizations we refer
to as \textit{views}. A central class named \texttt{Main} carries
out the initialization and coordination of the views. The Graphical
User Interface (GUI) is comprised of two main components, the \texttt{Setup
Panel} and the \texttt{Main Panel}. The \texttt{Main Panel} also serves
as a container for the views. The \texttt{Setup Panel} is the first
entity presented to the user when the application is opened in a web
browser. It serves mainly to specify the data types to be loaded and
the basic data features (spatial dimensionality, time resolution)
and visualization features such as the colors for each neuron population.
Parameters can be set manually or be loaded from configuration files,
for example specifying whether to load raw or preprocessed data. These
configuration files are JavaScript Object Notation\footnote{\href{http://www.json.org}{http://www.json.org}}
(JSON) files specifying the format of the loaded data, file names,
and preset values for the visualizations. After confirming the entered
information, the \texttt{Main.setup()} function extracts the entries
and provides them globally to the other components. The \texttt{Main
Panel} shown afterwards is used to load the input data files from
the local file system using JavaScript \texttt{FileReader}, which
includes setting up the internal data structure giving a coherent
access to the data to be visualized. The \texttt{Main.init()} function
then initializes the rendering loop, which is built into the web browser
and controls the rendering of the various views. The \texttt{Main.render()}
method is executed periodically by built-in functionality of the browser,
which is further used to synchronize the rendering of all views. This
is necessary as the rendering of an individual time step needs different
amount of time per view. For example, rendering a complex 3D scene
is slower (because it need more computational resources) than rendering
a less complex 2D plot. All views must have finished rendering before
the next rendering step is triggered. Each rendering call is triggered
by updating the global \texttt{timeStepIndex} through the browser.
The update of the \texttt{timeStepIndex} calls the rendering loop,
which redirects the rendering call to all views. For rendering the
data, all views access the loaded simulation data structure as part
of the\texttt{ Main} object. 

The \texttt{timeStepIndex} variable can be controlled in two modes.
First, it can be manually set by the user via a slider widget. The
slider widget enables the user to scroll manually along the time axis,
thus offering a simple navigation through the time series. Second,
the user can start an animation of the loaded data set by pressing
a start button. This triggers a periodic update of the \texttt{timeStepIndex}.
As both manual and periodical updates of the \texttt{timeStepIndex}
trigger the same downstream functionality for updating the views,
manual navigation through the time series can be combined with automatic
updates of the \texttt{timeStepIndex}. In case the animation is running,
manual intervention by the user overrides the periodic update of the
\texttt{timeStepIndex}, such that the shown data item (time step)
corresponds to that manually selected one. The animation is continued
from this manually selected time step.

All views offer view-specific selectors for visualization parameters.
These parameters are read in and used in each render call.

\subsection{2D and 3D view implementations}

This section describes the visualization algorithms used with the
different view implementations in VIOLA as introduced throughout the
\nameref{sec:results}. As the output data produced by the simulation
are scalars organized on a regular grid (related to the neuron's position),
visualization methods applicable to scalar data are employed as well
as standard chart types \citep{Hansen2011}. View 1 implements a standard
image plot in which the color of each spatial bin represents a spike-count
rate value. The display maps a binned measure of neuronal activity
on a color by using a lookup table ranging from black to white over
red and yellow (often referred to as \textquoteleft red hot\textquoteright{}
or \textquoteleft white hot\textquoteright{} lookup table). We selected
this lookup table as it is widely used and offers (if non-linear interpolated)
equidistant colors according to the CIE L{*}a{*}b{*} color space {[}EN
ISO 11664-4 from 1976{]}\footnote{\href{https://www.vis4.net/blog/posts/mastering-multi-hued-color-scales}{https://www.vis4.net/blog/posts/mastering-multi-hued-color-scales}}.
This color space is designed to represent equidistant colors according
to human perception: a color twice as light in CIE L{*}a{*}b{*} space
is also perceived twice as light by a human user \citep{Fairchild2013}.
Such a heat map \citep{Spence2001} facilitates the representation
of the spatial structure of the data.

In the 3D visualization of view 2, VIOLA also implements a geometrical
mapping of activity data to a cube's edge length, resulting in a cubic
mapping of the single scalar value representing activity at each time
step. The default scaling of the cube's edge length is such that the
cube's volume is proportional to the data value in each bin.

In view 3, the concept of a stacked bar chart supports a global perspective
on the simulated model \citep{Spence2001}. Individual populations
in the model are color-coded to be separable in the bar chart. An
additional line graph added on top of the bars shows the total spike-count
rate as reference to the global activity.

For view 4, in order to support the visual interpretation of time
series of spatially organized activity data, the 2D-organized activity
data (considering the neurons in one layer) are extracted along the
time axis resulting in a regularly structured 3D volume of scalar
values. Through contouring, partial sub-volumes with a certain minimum
threshold of activity get extracted and rendered as geometry. By means
of a selected iso-value $I_{\mathrm{th}}$ such geometry gets extracted
by applying the marching cubes algorithm for implicit volume rendering
\citep{Lorensen1987}. For extraction of the geometry, the algorithm
assumes that each data point of the data set is mapped onto a vertex
(corner point) of a regular 3D grid, which can be subdivided into
cells delimited by eight neighboring vertices each. Then, the algorithm
calculates for each vertex of a cell whether the associated data value
of the considered vertex lies inside or outside of the contour defined
by the iso-value $I_{\mathrm{th}}$ by comparing the data value with
the iso-value. If the data value of a vertex is smaller than the iso-value,
the vertex is assumed to lay inside of the contour. For each possible
combination of inside/outside states of the vertices of a cell, the
topology of the contour for each cell gets extracted from a table
by calculating a representative index. This table holds all possible
topological states of a cell, which are constructed under the assumption
that there are an infinite number of possibilities how a contour can
pass a cell (for more details, please refer to, for example \citet[Chap. 1]{Hansen2011}.
Finally, the exact position of the contour gets calculated by interpolation
along the cell's edges. 

Views 2, 3 and 4 all use the same color coding to identify the different
neuronal populations. The implementation of the algorithms and views
uses native JavaScript. The 2D rendering routine uses the HTML5 canvas
element. The browser rendering engine supports HTML5 and especially
the functionality of the canvas element, therefore no external libraries
are required. 3D renderings relied on the three.js\footnote{\href{https://threejs.org}{https://threejs.org}}
wrapper for WebGL content which is natively supported by the engine
of modern browsers. Node.js\footnote{\href{https://nodejs.org}{https://nodejs.org}}
facilitates the communication between views and the GUI for control. 

\subsection{Network description}

\label{sec:network_description}

The example network is based on an implementation of a random balanced
network \citep{Brunel2000} which is part of NEST as an example (\texttt{brunel\_alpha\_nest.py}
in NEST 2.12.0 by \citealp{Kunkel2017}). The model is expressed using
PyNEST \citep{Eppler2009} in Python\footnote{\href{http://www.python.org}{http://www.python.org}}.
The network consists of $N_{\mathrm{EX}}$ excitatory and $N_{\mathrm{IN}}$
inhibitory spiking point-neurons which are sparsely connected with
connection probability $c$. Neurons have fixed in-degrees of $cN_{\mathrm{EX}}$
excitatory and $cN_{\mathrm{IN}}$ inhibitory incoming synapses with
weights $g_{Y,\mathrm{EX}}J$ and $g_{Y,\mathrm{IN}}J$, respectively,
with $Y\in\{\mathrm{EX,IN}\}$. The integrate-and-fire model neurons
are connected using static, current-based synapses with an alpha-shaped
time course (NEST neuron model: \texttt{iaf\_psc\_alpha}). The intrinsic
neuron parameters are identical for both neuron types. In addition
to the recurrent connections, each neuron receives uncorrelated, external
excitatory input from a Poisson process of a fixed rate $\nu_{\mathrm{ext}}=\eta\nu_{\theta}$
, where $\eta$ denotes the external rate relative to the threshold
rate $\nu_{\theta}$ which is defined as $\nu_{\theta}=(V_{\theta}-E_{\mathrm{L}})C_{\mathrm{m}}/(\mathrm{exp}(1)J\tau_{\mathrm{m}}\tau_{\mathrm{s}})$.
The threshold rate is the hypothetical external rate needed to bring
the average membrane potential of a neuron to threshold $V_{\theta}$
(in the absence of an actual spiking mechanism). $E_{\mathrm{L}}$
denotes the resting potential, $C_{\mathrm{m}}$ the membrane capacitance,
$\tau_{\mathrm{m}}$ the membrane time constant and $\tau_{\mathrm{s}}$
the postsynaptic current time constant.

Unlike the original network implementation which has no spatial information,
we here place neurons randomly on a square 2D sheet with side lengths
$L$. The connection probability between a presynaptic neuron $j$
and postsynaptic neuron $i$ decays with increasing horizontal distance
$r_{ij}$ (using periodic boundary conditions) while we preserve the
in-degrees (number of incoming connections). A Gaussian-shaped profile
$p_{YX}\left(r_{ij}\right)$ is used with a standard deviation of
$\sigma_{YX}$ with $X,Y\in\{\mathrm{EX,IN}\}$. We use $\epsilon_{YX}$$\left(r_{ij}\right)$
to describe the distance-dependent connectivity profile assuming that
the in-degree is preserved. The transmission delay function $d_{YX}\left(r_{ij}\right)$
has a linear distance dependency with an offset $d_{YX}^{0}$ and
a conduction velocity $v_{YX}$.

In addition to the stationary external input to each population, the
network receives a spatially confined transient input with a duration
$t_{\mathrm{STIM}}$. The input is provided by a size $N_{\mathrm{STIM}}$
population of parrot neurons (NEST's \texttt{parrot\_neuron} devices),
positioned inside a circle of radius $R_{\mathrm{STIM}}$ around $(x,y)=(0,0)$.
Parrot neurons simply repeat input spike events as output spike events.
Each parrot neuron receives input from a Poisson process with a rate
expectation of $\nu_{\mathrm{STIM}}$ and connected to $K_{\mathrm{STIM}}$
neurons of the $\mathrm{EX}$ population inside a connection mask
radius $R$ from the parrot-neuron location. The Poisson input starts
at $T_{\mathrm{STIM}}$ and consequently the $\mathrm{STIM}$ units
become active after a delay of $d_{\mathrm{STIM}}$.

\prettyref{sec:network_description} summarizes the network description
with model and simulation parameters listed in \prettyref{tab:modelparams}A
and B.

The original parameters for the $\mathrm{EX}$-$\mathrm{IN}$ network
in the NEST example are modified for this VIOLA use case demonstration
bringing the network in a state with spatially confined network activity.
For this we increase the network's neuron count, reduce the ratio
of inhibitory to excitatory weights $g_{Y,\mathrm{IN}}$ and the membrane
capacitance $C_{\mathrm{m}}$, while the postsynaptic amplitude $J$
is increased. The parameter $J$ is originally defined in units of
$\mathrm{mV}$, but is here re-defined in units of $\mathrm{pA}$.
Finally, the fixed conduction delay is replaced by a distance-dependent
one.

The data sets result from simulations of duration $T_{\mathrm{sim}}$
with a temporal resolution of $dt$. We discard the startup transient
period $T_{\mathrm{trans}}$ and record all spike times from all neurons.
The unprocessed spike times together with the corresponding neuron
positions are considered as raw output. The temporal and spatial bin
sizes used for preprocessing, $\Delta t$ and $\Delta l$ respectively,
are given in \prettyref{tab:modelparams}C.

\subsection{LFP predictions}

\label{sec:LFP}

\emph{Generation of LFP-like data:} The local field potential (LFP)
is, due to its relative ease of measurement, a common measure of neuronal
activity \citep{Buzsaki2012,Einevoll2013}. The LFP is, in general,
assumed to reflect synaptic activity and correlations of a large number
of neurons in vicinity of the recording electrodes \citep{Kajikawa2011,Linden2011,Leski2013}.
For the purpose of demonstrating VIOLA's functionality, we synthesize
LFP signals from network activity assuming a linear network-population
spike to LFP relationship $H_{X}\equiv H_{X}(\vec{\Delta},\tau)$
derived using a biophysical model. In this relationship, $\vec{\Delta}$
denotes the displacement between the center of a spatial bin and an
electrode contact point $\gamma$ at $\mathbf{r}_{\gamma}$, and $\tau$
the time relative to a presynaptic spike event (``lag''). Assuming
linearity and homogeneous spike-LFP responses of individual presynaptic
neurons located within the same bin of width \foreignlanguage{english}{$\Delta l^{\phi}$}
indexed by $b=(l_{x}^{\phi},l_{y}^{\phi})$ (see \prettyref{sec:dataformats}),
the signal $\phi_{X}$ at one contact $\gamma$ of one population
$X$ is then given by 
\begin{equation}
\phi_{X}(\mathbf{r}_{\gamma},t)=\sum_{b}\left((\sum_{s}\delta(t_{b}^{s}))\ast H_{X}\right)(\mathbf{r}_{\gamma},t)~.\label{eq:phi_x}
\end{equation}
Here, the term $\sum_{s}\delta(t_{b}^{s})$ represents a series of
spike times $t_{b}^{s}$ of all presynaptic neurons in a bin $b$
where $\delta$ denotes the Dirac delta function, and $\ast$ a convolution.
As contributions of different populations $X$ sum linearly, the total
signal at each contact is 
\begin{equation}
\phi(\mathbf{r}_{\gamma},t)=\sum_{X}\phi_{X}(\mathbf{r}_{\gamma},t)~.
\end{equation}

Point-like neurons (as used in our network model) can not generate
an extracellular potential, as all in- and outgoing currents sum to
zero at the point's location (due to conservation of charge). As in
\citet{Hagen2016} we assume that spatially extended (morphologically
detailed) neurons and corresponding multicompartment models in combination
with an electrostatic forward model are required to compute a biophysically
meaningful LFP signal. To compute the LFP, we here derive for each
presynaptic population $X\in\{\mathrm{EX,IN,STIM}\}$ the phenomenological
mapping $H_{X}(\vec{\Delta},\tau)$ between a presynaptic spike event
time $t_{b}^{s}$ occurring in a spatial bin indexed by $b$ to the
extracellular potential.

\emph{Measurement sites:} The electrode contact point locations are
defined at the center of each spatial bin as $\mathbf{r}_{\gamma}=((l_{x}^{\phi}+1/2)\Delta l^{\phi}-L/2,(l_{y}^{\phi}+1/2)\Delta l^{\phi}-L/2,0)$.

\emph{Multicompartment model:} We define a ball-and-stick type multicompartment
model neuron with morphological features and passive parameters derived
from the network's LIF neuron description (membrane capacitance $C_{\mathrm{m}}$,
membrane time constant $\tau_{\mathrm{m}}$, passive leak reversal
potential $E_{\mathrm{L}}$). Assuming a homogeneous specific membrane
capacitance $c_{\mathrm{m}}$ (capacitance per membrane area) and
axial resistivity $r_{\mathrm{a}}$ (resistance times length unit),
we choose the dendritic stick length $L_{\mathrm{dend}}$ and radius
$r_{\mathrm{dend}}$ as follows: To preserve the total capacity of
the point neuron (and equivalent surface area), we compute the corresponding
soma radius as $r_{\mathrm{soma}}=\sqrt{\frac{C_{\mathrm{m}}}{4\pi c_{\mathrm{m}}}-\frac{r_{\mathrm{dend}}L_{\mathrm{dend}}}{2}}$.
We define the passive leak conductivity as $g_{\mathrm{L}}=c_{\mathrm{m}}/\tau_{\mathrm{m}}$
and leak reversal potential as $E_{\mathrm{L}}$. For these calculations
we choose $c_{\mathrm{m}}$, $r_{\mathrm{a}}$, $L_{\mathrm{dend}}$
and $r_{\mathrm{dend}}$ values as given in \prettyref{tab:LFP},
resulting in $r_{\mathrm{soma}}\approx\unit[13.1]{\mu m}$. The compact
ball-like soma is treated as a single segment, while the elongated
dendrite is split into $n_{\mathrm{dend}}=11$ segments of equal length.
The center of the soma segment is set to $\mathbf{r}=(0,0,0)$, and
the dendritic stick is aligned in the positive direction along the
vertical $z-$axis.

\emph{Synapse model}: For LFP predictions we use the same current-based
synapse model as in the network, defining the postsynaptic input current
of a single presynaptic spike event as $I_{ij}(t)=J_{YX}\cdot(t-t_{j}^{s}-d_{ij})/\tau_{\mathrm{syn}}\exp(1-(t-t_{j}^{s}-d_{ij})/\tau_{\mathrm{syn}})\Theta(t-t_{j}^{s}-d_{ij})$,
where $J_{YX}$ denotes the connection-specific postsynaptic current
amplitude as in the network, $t_{j}^{s}$ the presynaptic spike time,
$d_{ij}=d_{YX}(r_{ij})$ the conduction delay between presynaptic
cell $j$ and postsynaptic cell $i$ and $\Theta$ the Heaviside step
function. As we initially ignore delays and network spike times we
set $d_{ij}=0$ and $t_{j}^{s}=\tau^{s}$.

\emph{Synaptic connectivity}: For outgoing connections of the excitatory
populations $X\in\{\mathrm{EX,STIM}\}$ we distribute synaptic input
currents evenly along the entire length of the dendritic stick, while
for outgoing connections of the inhibitory population $X=\mathrm{IN}$
all synaptic input currents are assumed to be evenly distributed on
the ball-like soma.

\emph{Electrostatic forward model:} As described in detail in \citet{Linden2014},
we assume an extracellular conductive medium that is linear (frequency
independent), isotropic (identical in all directions), homogeneous
(identical in all positions) and ohmic (linear relationship between
current density and electric potential), as represented by the scalar
conductivity $\sigma_{\mathrm{e}}$ (cf. \prettyref{tab:LFP} for
values). From the linearity of Maxwell's equations, contributions
to the extracellular potential from different current sources sum
linearly. Here, these current sources are transmembrane currents (summed
over resistive, capacitive and synaptic currents). In the presently
used volume conduction theory, the electric potential in location
$\mathbf{r}_{\gamma}$ from a point current with magnitude $I(t)$
in location $\mathbf{r}_{0}$ is 
\begin{equation}
\phi_{\mathrm{point}}(\mathbf{r}_{\gamma},t)=\frac{1}{4\pi\sigma_{\mathrm{e}}}\frac{I(t)}{|\mathbf{r}_{\gamma}-\mathbf{r}_{0}|}~.
\end{equation}
This relation is also valid for a sphere current source (i.e., our
ball soma) centered at $\mathbf{r}_{0}$ with total transmembrane
current $I_{\mathrm{m},\mathrm{soma}}$ and radius $r_{\mathrm{sphere}}$
when $|\mathbf{r}_{\gamma}-\mathbf{r}_{0}|\geq r_{\mathrm{sphere}}$.
Thus 
\begin{equation}
\phi_{\mathrm{soma}}(\mathbf{r}_{\gamma},t)=\frac{1}{4\pi\sigma_{\mathrm{e}}}\frac{I_{\mathrm{m,soma}}(t)}{|\mathbf{r}_{\gamma}-\mathbf{r}_{\mathrm{soma}}|}~.
\end{equation}
The elongated dendritic segments are treated as \textquoteleft line
sources\textquoteright , obtained by integrating the point-source
formula along the central axis of the segments \citep{Holt1999,Linden2014}:
\begin{equation}
\phi_{\mathrm{dend}}(\mathbf{r}_{\gamma},t)=\frac{1}{4\pi\sigma_{\mathrm{e}}}\sum_{u=1}^{n_{\mathrm{dend}}}I_{\mathrm{m},u}(t)\int\frac{d\mathbf{r}_{u}}{|\mathbf{r}_{\gamma}-\mathbf{r}_{u}|}~.
\end{equation}
The total extracellular potential from somatic and dendritic sources
is then 
\begin{equation}
\phi(\mathbf{r}_{\gamma},t)=\phi_{\mathrm{soma}}(\mathbf{r}_{\gamma},t)+\phi_{\mathrm{dend}}(\mathbf{r}_{\gamma},t)~.
\end{equation}

Our calculations of extracellular potentials rely on the Python package
\texttt{LFPy}\footnote{\href{http://lfpy.github.io}{http://lfpy.github.io}}
(\citet{Linden2014,Hagen2018}). The tool implements the above forward-model
formalism for extracellular potentials, and uses the NEURON simulation
environment \citep{Carnevale2006} to compute transmembrane currents
$I_{\mathrm{m}}(t)$ of multicompartment neuron models. As singularities
may occur in the limit $|\mathbf{r}_{\gamma}-\mathbf{r}_{u}|\rightarrow0$,
the minimum distance between sources and measurement locations was
set equal to the somatic or dendritic segment radius.

\emph{Prediction of spike-LFP relationship:} We here describe the
calculation of the linear spike-LFP relationships $H_{X}(\vec{\Delta},\tau)$
which we use to construct an LFP-like signal from spatially binned
network activity. While \citet{Hagen2016} present a hybrid scheme
to compute extracellular potentials from point-neuron network activity,
and incorporated the biophysics-based forward model summarized above,
this hybrid scheme is not adapted to laminar point-neuron networks
with distance-dependent connections. We therefore construct a simpler
and numerically much less demanding method inspired by the hybrid
scheme, that still encompasses the governing biophysics underlying
the generation of extracellular potentials and accounts for the laminar
structure and distance-dependent connectivity of our network.

In this simplified model, we ignore heterogeneity in spike-LFP responses
$H_{i}$, of individual presynaptic cells $i\in X$ located within
a spatial bin $b$, i.e., $H_{X}\equiv\langle H_{i}\rangle$. $H_{i}$
corresponds to the extracellular potential resulting of synaptic activation
of postsynaptic populations of cells $j\in Y$ from a spike in cell
$i$ at time $\tau=0$. We also assume that $H_{X}$ is invariable
across presynaptic bins, and encompasses the overall distance-dependent
connection probabilities and connection delays in the network.

The calculation of $H_{X}(\vec{\Delta},\tau)$ involves a number of
steps. We first estimate the spatially averaged extracellular potential
$\varphi_{j}(\vec{\Delta},\tau)$ resulting from a single synapse
activation at a time $\tau^{s}$ of the ball and stick neuron positioned
at the center of a reference bin, for excitatory and inhibitory input.
Electrode contact point locations $\mathbf{r}_{\gamma}$ are defined
at the centers of each square spatial bin indexed $b$ (see above).
With rotational symmetry around the $z-$axis and periodic (torus)
connectivity of the network, we compute extracellular potentials at
the unique subset of bin center-to-center distances $r\subset\{|\vec{\Delta}|\}$
up to the maximum distance $\sqrt{2L^{2}}$, where $L$ denotes the
side length of the network layers, and $\{|\vec{\Delta}|\}$ the complete
set of center-to-center displacement vector lengths from reference
bin to all spatial bins. We utilize built-in functionality in \texttt{LFPy}
to perform spatial averaging (cf. Eq. 6 in \citet{Linden2014}), assuming
square contact points parallel to the horizontal $xy-$plane with
side lengths equal to the bin width \foreignlanguage{english}{$\Delta l^{\phi}$}.
In a following step we compute the average out-degree (number of outgoing
connections of neuron $i$) $K_{X}=\sum_{Y}N_{Y}c$ for $X\in\{\text{EX, IN}\}$,
where $c$ denotes the overall connection probability between $X$
and $Y$ (cf. \prettyref{tab:modelparams} which also gives $K_{\mathrm{STIM}}$
as a fixed parameter). With the distance-dependent connectivity $\epsilon_{YX}(r)$
used for each presynaptic population and out-degree $K_{X}$ we compute
the number of activated synapses (denoted by $K_{r}$) in each spatial
bin at a distance $r$ from the reference bin (including $r=0$) by
evaluating $p_{YX}(r)$ at the bin center points. The average connection
delays from the reference bin to other bins are approximated as $d_{YX}(r)=v_{YX}+w_{YX}r$,
where $v_{YX}$ denotes a constant delay offset and $w_{YX}$ the
inverse propagation speed of action potentials in the network, as
summarized in \prettyref{tab:modelparams} With the elements of these
steps in place (single-synapse LFP responses across bins, bin-wise
number of activated synapses and delays), we construct $H_{X}(\vec{\delta},\tau)$
as function of $r$ as: 
\begin{equation}
H_{X}(\vec{\Delta},\tau)=\sum_{r\in\{|\vec{\Delta}|\}}K_{r}\cdot(\delta(d_{YX}(r))\ast\varphi_{j})(\vec{\Delta},\tau)~,
\end{equation}
where $\delta(\cdot)$ denotes the Dirac delta function. Note that
we sum over all elements $r$ in $\{|\vec{\Delta}|\}$.

\emph{LFP output:} Each $H_{X}$ is calculated at a spatial resolution
\foreignlanguage{english}{$\Delta l^{\phi}$} and temporal resolution
of $dt$ (as in the network, cf. \prettyref{tab:modelparams}) for
a total duration of $2\tau^{s}$, with synapse activation time at
time $\tau^{s}$. An identical spatial and temporal binning resolution
is also used for spike events entering in \prettyref{eq:phi_x}. The
spike rates in each bin are filtered by a length $\Delta t$ normalized
boxcar filter using the \texttt{scipy.signal.lfilter} method prior
to the convolution with the corresponding LFP kernel. Otherwise a
temporal shift between the spatiotemporally binned spiking data (cf.
\prettyref{sec:dataformats}) and the downsampled LFP in the visualization
occurs. Discrete convolutions are incorporated using \texttt{numpy.convolve}
and \texttt{scipy.signal.convolve2d} methods in Python. The final
LFP signals are low-pass filtered and downsampled to the time resolution
$\Delta t$ of our preprocessed network output as described in \citet{Hagen2016}
in order to simultaneously show both datasets in VIOLA. Output is
stored in a pure-text format as described in \prettyref{sec:dataformats}.

\prettyref{tab:LFP} summarizes the parameter values for the LFP predictions.

\subsection{Software summary}

All source codes of the tool VIOLA, the example network model and
the processing of model output are hosted at \href{https://github.com/HBPVIS/VIOLA}{https://github.com/HBPVIS/VIOLA}
(SHA:ca2f3c5). We simulated the example network (\texttt{topo\_brunel\_alpha\_nest.py})
with NEST v2.12.0 and Python v2.7.11. Further processing and plotting
of Figures \ref{fig:network_sketch} and \ref{fig:raster} (\texttt{nest\_preprocessing.py})
also relied on Python with numpy v1.10.4, SciPy v0.17.0, and matplotlib
v1.5.1. LFP signals (\texttt{fake\_LFP\_signal.py}) were computed
using NEURON v7.5 and LFPy from \href{http://lfpy.github.io}{http://lfpy.github.io}
(SHA:5673a6). We visualized the neuronal activity with VIOLA using
the Google Chrome browser, version 58.0.3029.110 (64-bit). VIOLA used
JavaScript V8 5.8.283.38 with the 3D library three.js of revision
87, including WebGL and HTML5 build in the browser and Node.js v4.8.3.
For colors, VIOLA used Chroma.js in the version 1.3.4.

Screenshots from VIOLA for the other figures were taken with Kazam-``NCC-80102''
v1.4.5, and combined in Microsoft PowerPoint 2013.

\section{Discussion and Outlook}

\label{sec:discussion}

The present study introduces four 2D and 3D visualization concepts,
or views, for the interactive visual analysis of the activity of spiking
neuronal network simulations, and a reference implementation for these
views named VIOLA (VIsualization of Layer Activity). VIOLA is an interactive
web-technology based visualization tool designed to fit in between
simulations and subsequent in-depth data analysis, and exemplifies
key concepts of the information-seeking mantra by \citet{Shneiderman1996}
and the paradigm of coordinated multiple views \citep{Wang2000}.
The main application areas are the rapid validation of simulation
results and the exploration of spatiotemporally resolved data prior
to further quantitative analyses. As a use case, we demonstrate the
usefulness of the tool with output from a simulation of a layered
spiking point-neuron network model that incorporates distance-dependent
connectivity. The use case shows that we can examine a perturbation
of ongoing network activity caused by a temporally and spatially confined
stimulus. The duration and the spatial spread of the event are quickly
assessed with the help of multiple simultaneously displayed views.

In contrast to other visualization tools for simulated network output,
for example VisNEST \citep{Nowke2013,Nowke2015}, SNN3DViewer \citep{Kasinski2009},
ViSimpl \citep{Galindo2016}, and Geppetto or more generic multi-view
tools like GLUE, the interactive JavaScript- and WebGLbased visualization
integrates data analysis methods in a web application, thereby achieving
mobility and deployability. Our approach builds on visualization concepts
known from the literature for data of similar structure (reviewed
in the introduction), but advances the concepts and adds interactivity
and animation. For example, views 1 and 2 compare to series of snap
shots (as in \citealt{Mehring2003,Yger2011,Voges2012,Keane2015}),
but are here enhanced by the possibilities to show raw or preprocessed
data, to specify visualization parameters interactively, and to provide
a 3D and temporally animated view on the multi-dimensional data. View
4 presents a new concept combining 2D spatial and temporal resolution
of multiple neuron populations, all shown simultaneously. This data
representation delivers a wealth of information, but, to circumvent
occlusion and instead expose interesting features of the data, it
relies on interactive usage. The code of the reference implementation
is open source and available in a public repository (\href{https://github.com/HBPVIS/VIOLA}{https://github.com/HBPVIS/VIOLA})
together with the revision history and documentation. The present
work uses the simulation code NEST to generate the data but the VIOLA
implementation is completely independent of the former. The JavaScript
code defines a standalone application (accessible at \href{http://hbpvis.github.io/VIOLA}{http://hbpvis.github.io/VIOLA})
and interpretable by the browser running on the client device. In
the last decade, JavaScript-based visualization got more and more
versatile especially fostered by the introduction of HTML5 and its
canvas environment. Furthermore, the development of WebGL enables
the access to GPU-accelerated 3D rendering in the browser. Beside
limitations regarding memory and access to low-level program control
(as needed for controlled use of multi-threading), JavaScript offers
the opportunity for simple deployment and handling of external libraries
and dependencies. Unfortunately, JavaScript-based implementations
tend to fail on one or the other browsers as browsers still differ
in their interpretation of JavaScript and in the degree of following
the HTML standard. Nevertheless, most browsers are free to use and
usable on most operating systems. Therefore, this work explores the
radical decision to use web-based technology to offer an easy-to-deploy
tool for the visualization of dynamic simulation data. As a consequence,
software development and deployment are integrated with minimal effort
and no computational resources are required on the server: researchers
immediately profit from progress on the development platform. Furthermore,
due to the web-technology and the minimal requirements on the client,
web portals can embed the application as a visualization backend;
a prerequisite for the idea to create centralized ICT infrastructure
for neuroscience. One such portal is currently being developed by
the European Human Brain Project, named the HBP Collaboratory\footnote{\href{https://collab.humanbrainproject.eu}{https://collab.humanbrainproject.eu}}.
Another ongoing effort is the Neuroscience Gateway\footnote{\href{http://www.nsgportal.org}{http://www.nsgportal.org}}
\citep{Sivagnanam2013}. Online embedding opens the possibility to
accompany interactive visualization with server-side preprocessing
steps and a database integration, in particular for simulation output
being generated on the portal itself. This advances the goal of the
HBP Collaboratory to provide a fully digitized workflow from data
representation over model construction and simulation to model validation
\citep{Senk2017}. We argue that interactive visual analysis of simulated
data is an obvious feature of a collaboratory, in addition to non-interactive
script-based plotting relying for example on \texttt{matplotlib}. 

The reference implementation accesses the file system of the host
machine to load data. This is not recommended for web applications
for security reasons. If data processing and storage were handled
on the server-side, SQL-like database queries could restrict communication
to only the data needed for the different view instances. Communication
does not have to be limited to the raw data. Binning operations similar
to those performed in our preprocessing steps can be handled by the
database in a straightforward manner, and could also be performed
in parallel. The data format HDF\footnote{\href{https://www.hdfgroup.org}{https://www.hdfgroup.org}}
would also be an option to store and access large amounts of raw and
preprocessed data with improved performance in terms of speed and
compactness compared to the currently used text format.

Inherent in interactive visualization is the problem of reproducibility.
The raw data are insufficient to reproduce the visuals, only in combination
with the full collection of GUI parameters adjusted by the researchers
is the data set complete. In the same way as experimental and simulated
data need to be enriched with metadata in order to uniquely specify
their origin and enable reuse \citep{ZehlJailletStoewerEtAl2016},
the visuals need to be enriched with the parameters of their creation.
This new type of metadata could be stored in a database.

The JavaScript implementation imposes other shortcomings. Prominent
is its limited capability for numerical analysis. While the \texttt{math.js}\footnote{\href{http://mathjs.org}{http://mathjs.org}}
library provides a number of basic math functions and support for
symbolic operations, complex numbers and arrays (matrices), the JavaScript
libraries are not comparable to the Scientific Python stack (SciPy\footnote{\href{https://www.scipy.org}{https://www.scipy.org}})
which provides an ecosystem of fundamental tools and methods encountered
in mathematics, engineering and science. VIOLA implements the function
computing the spatial correlation of neuronal activity from scratch
(not shown). This approach has two conceptual weaknesses. First, the
speed and accuracy of such functions are hampered by the fact that
there is little native support for advanced mathematical operations,
like the Fast Fourier Transform (FFT). Second, there is no separation
between the code carrying out the statistical analysis and the code
performing the visualization. This cuts visualization off from the
rich set of analysis tools developed by the community and their reliable
implementations, for example as collected in the Elephant package\footnote{\href{http://elephant.readthedocs.io}{http://elephant.readthedocs.io}}.
Future work needs to disentangle numerics from visualization code
as separate building blocks in a visual analysis workflow.

As VIOLA's main focus lies on responsive interactive visualization,
the reference implementation uses WebGL for all views. Prior tests
exposed the low efficiency of the Document Object Model (DOM) as used
in Scalable Vector Graphics (SVG) based visualization libraries such
as d3.js as well as its high memory consumption. This led to the decision
to the sole use of WebGL rendering, which has the limitation that
external tools are required for generating screen shots and screen
casts; vector graphics can neither be recorded nor exported. For the
2D views, an additional implementation based on the HTML 5 support
of SVG graphics can be added and used for the export of vector-based
image material. For extracting vector-based material from the 3D views,
WebGL and its access to the underlying rendering pipeline can be facilitated.
The 3D scene can be exported to be viewed in other 3D programs. To
this end, three.js (as used in the reference implementation) offers
export functionality for Wavefront OBJ file format, one standard for
3D content. The alternative is to extract the rendered scene prior
to rasterization and use these data to generate a SVG or postscript-based
representation similar to the operation of the C library gl2ps\footnote{\href{http://www.geuz.org/gl2ps}{http://www.geuz.org/gl2ps}}.
Nevertheless, any export mechanism needs to facilitate means of reproducibility.
In particular metadata such as simulation and visualization parameters,
time stamps, viewpoint angle and position etc. need to be bundled
with the raw visualizations. As direct file writes may not be possible
in a client side JavaScript application, one solution is server-based
rendering and storage based on visualization parameters being communicated
from the client back to the server. The resulting server-side rendered
images are then stored as provenance information. Another option
for reproducible visualization outcome is to only store the previously
mentioned visualization parameters in the database, such that the
client-side visualization application can be set back into the original
captured state. If these parameters are captured over a longer period,
the resulting data can ease the regeneration of content for demonstration
purposes or a post-hoc video rendering.

While we here develop our arguments along model data, the different
views and the reference implementation is equally suited for the exploration
of experimental data. Our model network describes a neuronal layer
covering a $\unit[4\times4]{mm^{2}}$ patch of cortical tissue. Electrophysiological
measurements of neuronal activity with the Utah multi-electrode array
from Blackrock Microsystems sample both spiking activity of individual
cells and population LFPs across near $\unit[4\times4]{mm^{2}}$ of
cortex \citep{Milekovic2015,Torre2016,Denker2017}. Other multi-electrode
arrays are used for \textit{\emph{in vitro}} experimentation on neural
tissue or cell cultures \citep{Massobrio2015}. No changes to the
reference implementation are required for the processing of these
data. Other measurement modalities are of interest as well. One
common experimental method is Ca$^{2+}$ imaging which may infer changes
in intracellular {[}Ca$^{2+}${]} of neurons in superficial \citep{Grienberger2012}
and deep layers \citep{Ouzounov2017}, while another method is voltage-sensitive
dye imaging (VSDi) that measures membrane-voltage time-derivatives
in surface-proximal tissues \citep{Chemla2010}. With modifications
to existing views or new view implementations, VIOLA can also represent
this type of spatiotemporally resolved data. In particular the 3D
visualization types incorporated in the present views 2 and 4 are
well suited to represent the changes in intracellular $\mathrm{Ca^{2+}}$
ion concentrations across different cell bodies from 2- and 3-photon
volumetric $\mathrm{Ca^{2+}}$ imaging in neural tissue. Within view
2 the visual representations of each cell's concentration can be set
to a depth and position in the horizontal plane according to its image
stack position in the raw imaging data. Units with baseline $\mathrm{Ca^{2+}}$
concentrations may then be hidden, and increasing levels can be visualized
by scaling the box sizes as we have demonstrated with spike-rate data.
A view similar to view 4 could show time-varying ion-concentrations
of individual units as 3D tube plots where the tube diameter at a
given time is proportional to a unit's $\mathrm{Ca^{2+}}$ concentration.
As VSDi imaging data (typically) lacks depth-information, color-image
plotting can be applied similar to what we utilize here to show LFPs
in our 3D view implementations. In addition, the multi-view aspect
of visualization enables the combination of spatial representations
with more abstract non-spatial representations of neuronal activity,
as reviewed in the introduction.

The concepts developed here advance the visual exploration of data
from cortical networks at cellular resolution. If the reference implementation
finds more widespread interest it can be further developed by a community
driven approach as all requirements like a proper licensing and a
suitable development platform are in place, the primary purpose, however,
is to serve as a living supplement to this publication. Creating a
common web portal for the collaboration of neuroscientists is a central
long-term goal of the Human Brain Project. In this endeavor our study
contributes knowledge on how a user interface for visual exploration
needs to be designed and on the proper layout of the software stack
at the troubled transition point between data processing and visualization.

\section*{Author Contributions}

JS, EH, CC, TK, MD, BW conceived and designed the study. JS, EH, CC,
BW designed the various types of visualizations. CC implemented the
first version of VIOLA and JS co-developed the tool. JS, EH implemented
simulation, preprocessing and analysis code for the example network.
EH incorporated LFP predictions. JS ran all simulations and created
the figures. JS wrote the first version of the paper. EH, BW, MD,
CC, TK co-wrote the paper.

\section*{Acknowledgments}

This project received funding from the European Union's Horizon 2020
research and innovation programme under grant agreement No. 720270
(HBP SGA1), \foreignlanguage{english}{the European Union Seventh Framework
Programme {[}FP7/2007-2013{]} under grant agreement No. 604102 (Human
Brain Project, HBP),} the Helmholtz Association through the Helmholtz
Portfolio Theme ``Supercomputing and Modeling for the Human Brain'',
the Research Council of Norway (NFR) through COBRA. 

\FloatBarrier

\begin{table}[t]
\begin{raggedright}
\begin{tabular}{|@{\hspace*{1mm}}p{0.2\linewidth}@{}|@{\hspace*{1mm}}p{0.75\linewidth}|}
\hline 
\multicolumn{2}{|>{\color{white}\columncolor{black}}l|}{\textbf{A: Model summary}}\tabularnewline
\hline 
\textbf{Populations} & Three: excitatory $\mathrm{EX}$, inhibitory $\mathrm{IN}$, external
stimulus $\mathrm{STIM}$\tabularnewline
\hline 
\textbf{Topology} & $\mathrm{EX}$/$\mathrm{IN}$: random neuron positions on square domain
of size $L\times L$; $\mathrm{STIM}$: random neuron positions inside
a circle with radius $R_{\mathrm{STIM}}$ at the center of the domain;
periodic boundary conditions\tabularnewline
\hline 
\textbf{Connectivity} & Random ($\mathrm{EX}$/$\mathrm{IN}$: convergent, fixed in-degree;
$\mathrm{STIM}$: divergent, fixed out-degree) connections described
by distance-dependent probability kernels and cut-off masks\tabularnewline
\hline 
\textbf{Neuron model} & $\mathrm{EX}$/$\mathrm{IN}$: leaky integrate-and-fire (LIF), fixed
threshold, absolute refractory time; $\mathrm{STIM}$: parrot\tabularnewline
\hline 
\textbf{Synapse model} & Static weights, $\mathrm{EX}$/$\mathrm{IN}$: alpha-shaped postsynaptic
currents, distance-dependent delays\tabularnewline
\hline 
\textbf{Input} & Independent fixed-rate Poisson spike trains to all neurons\tabularnewline
\hline 
\textbf{Measurement} & Spike activity\tabularnewline
\hline 
\end{tabular}
\par\end{raggedright}
\begin{raggedright}
\begin{tabular}{|@{\hspace*{1mm}}p{0.2\linewidth}@{}|@{\hspace*{1mm}}p{0.75\linewidth}|}
\hline 
\multicolumn{2}{|>{\color{white}\columncolor{black}}l|}{\textbf{B: Network model}}\tabularnewline
\hline 
\textbf{Subthreshold dynamics} & $\mathrm{EX}$/$\mathrm{IN}$:\newline If $t>t^{*}+\tau_{\mathrm{ref}}$\begin{itemize}\item[]$\frac{\mathrm{d}V}{\mathrm{d}t}=-\frac{V-E_{L}}{\tau_{\mathrm{m}}}+\frac{I_{\mathrm{syn}}\left(t\right)}{C_{\mathrm{m}}}$\item[]$I_{\mathrm{syn}}\left(t\right)=\sum_{j}J_{j}\alpha\left(t-t_{j}^{*}-d_{j}\right)$\item[]
with connection strength $J_{j}$, presynaptic spike time $t_{j}^{*}$
and conduction delay $d_{j}$ \item[]$\alpha\left(t\right)=\mathrm{\frac{t}{\tau_{\mathrm{s}}}e}^{1-t/\tau_{\mathrm{s}}}\Theta\left(t\right)$
with Heaviside function $\Theta$\end{itemize} else \begin{itemize}
\item[] $V\left(t\right)=V_{\mathrm{reset}}$\end{itemize}\tabularnewline
\hline 
\textbf{Spiking} & If $V\left(t-\right)<V_{\theta}\wedge V\left(t+\right)\geq V_{\theta}$\begin{enumerate}\item
set $t^{*}=t$ \item emit spike with timestamp $t^{*}$ \item reset
$V\left(t\right)=V_{\mathrm{reset}}$\end{enumerate}\tabularnewline
\hline 
\textbf{Distance-dependent connectivity} & Neuronal units $j\in X$ at location $\left(x_{j},y_{j}\right)$ and
$i\in Y$ at $\left(x_{i},y_{i}\right)$ in pre- and postsynaptic
populations $X$ and $Y$, respectively.\newline Distance between
units $i$ and $j$:\begin{itemize} \item[]$r_{ij}=\sqrt{\left(x_{i}-x_{j}\right)^{2}+\left(y_{i}-y_{j}\right)^{2}}$\end{itemize}
Gaussian kernel for connection probability: \begin{itemize} \item[]
$p_{YX}(r_{ij})=\mathrm{e}^{-r_{ij}^{2}/2\sigma_{YX}^{2}}$ \end{itemize}
$R$ is the radius of a cut-off mask.\newline Transmission delay
function: \begin{itemize} \item[] $d_{YX}(r_{ij})=d_{YX}^{0}+r_{ij}/v_{YX}$\end{itemize}\tabularnewline
\hline 
\end{tabular}
\par\end{raggedright}
\caption{Description of the network model following the guidelines of \citet{Nordlie2009}.
\label{tab:netwdescription}}
\end{table}

\begin{table}[t]
\begin{tabular}{|@{\hspace*{1mm}}p{0.1\linewidth}@{}|@{\hspace*{1mm}}p{0.2\linewidth}|@{\hspace*{1mm}}p{0.65\linewidth}|}
\hline 
\multicolumn{3}{|>{\color{white}\columncolor{black}}l|}{\textbf{A: Global simulation parameters}}\tabularnewline
\hline 
\textbf{Symbol} & \textbf{Value} & \textbf{Description}\tabularnewline
\hline 
$T_{\mathrm{sim}}$ & $\unit[1,500]{ms}$ & Simulation duration\tabularnewline
$dt$ & $\unit[0.1]{ms}$ & Temporal resolution\tabularnewline
$T_{\mathrm{trans}}$ & $\unit[500]{ms}$ & Startup transient\tabularnewline
$T_{\mathrm{STIM}}$ & $\unit[999]{ms}$ & Start time of Poisson input to $\mathrm{STIM}$\tabularnewline
$t_{\mathrm{STIM}}$ & $\unit[50]{ms}$ & Duration of $\mathrm{STIM}$ onset\tabularnewline
\hline 
\end{tabular}

\begin{tabular}{|@{\hspace*{1mm}}p{0.1\linewidth}@{}|@{\hspace*{1mm}}p{0.2\linewidth}|@{\hspace*{1mm}}p{0.65\linewidth}|}
\hline 
\multicolumn{3}{|>{\color{white}\columncolor{black}}l|}{\textbf{B: Point-neuron network}}\tabularnewline
\hline 
\multicolumn{3}{|>{\color{black}\columncolor{lightgray}}c|}{\textbf{Populations and external input}}\tabularnewline
\hline 
\textbf{Symbol} & \textbf{Value} & \textbf{Description}\tabularnewline
\hline 
$X$ & $\mathrm{EX}$, $\mathrm{IN}$, $\mathrm{STIM}$ & Name\tabularnewline
$N_{X}$ &  & Population size:\tabularnewline
 & $20,000$ & $X=\mathrm{EX}$\tabularnewline
 & $5,000$ & $X=\mathrm{IN}$\tabularnewline
 & $975$ & $X=\mathrm{STIM}$\tabularnewline
$L$ & $\unit[4]{mm}$ & Extent length\tabularnewline
$\eta$ & $2$ & External rate relative to threshold rate for $X\in\left\{ \mathrm{EX},\mathrm{IN}\right\} $\tabularnewline
$R_{\mathrm{STIM}}$ & $\unit[0.5]{mm}$ & Radius of circle around $\left(0,0\right)$ for locations of $\mathrm{STIM}$\tabularnewline
$\nu_{\mathrm{STIM}}$ & $\unit[300]{Hz}$ & External rate to each $\mathrm{STIM}$ neuron\tabularnewline
\hline 
\multicolumn{3}{|>{\color{black}\columncolor{lightgray}}c|}{\textbf{Connection Parameters}}\tabularnewline
\hline 
\textbf{Symbol} & \textbf{Value} & \textbf{Description}\tabularnewline
\hline 
$c$ & $0.1$ & Connection probability for recurrent connections between $\mathrm{EX}$
and $\mathrm{IN}$\tabularnewline
$J$ & $\unit[40]{pA}$ & Reference synaptic strength. All synapse weights are measured in units
of $J$.\tabularnewline
$g_{YX}$ &  & Relative synaptic strengths:\tabularnewline
 & $1$ & $\ensuremath{X=\mathrm{EX},\,Y\in\{\mathrm{EX,IN}\}}$\tabularnewline
 & $-4.5$ & $\ensuremath{X=\mathrm{IN},\,Y\in\{\mathrm{EX,IN}\}}$\tabularnewline
 & $1$ & $X=\mathrm{STIM},\,Y=\mathrm{EX}$\tabularnewline
$R$ & $\unit[0.1]{mm}$ & Radius of cut-off mask for $\mathrm{X=STIM,\,Y=EX}$\tabularnewline
$K_{\mathrm{STIM}}$ & $300$ & Number of connections per $\mathrm{STIM}$ neuron\tabularnewline
$\sigma_{YX}$ &  & Standard deviation of Gaussian kernel:\tabularnewline
 & $\unit[0.3]{mm}$ & $X,Y\in\left\{ \mathrm{EX,IN}\right\} $\tabularnewline
$d_{YX}^{0}$ &  & Delay offset:\tabularnewline
 & $\unit[0.5]{ms}$ & $X,Y\in\left\{ \mathrm{EX,IN}\right\} $\tabularnewline
 & $\unit[0.5]{ms}$ & $X=\mathrm{STIM},\,Y=\mathrm{EX}$\tabularnewline
$v_{YX}$ &  & Conduction velocity:\tabularnewline
 & $\unit[2]{m/s}$ & $X,Y\in\left\{ \mathrm{EX,IN}\right\} $\tabularnewline
 & $-$ & $X=\mathrm{STIM},\,Y=\mathrm{EX}$\tabularnewline
$d_{\mathrm{STIM}}$ & $\unit[0.5]{ms}$ & Delay from Poisson input to $\mathrm{STIM}$\tabularnewline
\hline 
\multicolumn{3}{|>{\color{black}\columncolor{lightgray}}c|}{\textbf{Neuron model}}\tabularnewline
\hline 
\textbf{Symbol} & \textbf{Value} & \textbf{Description}\tabularnewline
\hline 
$C_{\mathrm{m}}$ & $\unit[100]{pF}$ & Membrane capacitance \tabularnewline
$\tau_{\mathrm{m}}$ & $\unit[20]{ms}$ & Membrane time constant \tabularnewline
$E_{\mathrm{L}}$ & $\unit[0]{mV}$ & Resting potential \tabularnewline
$V_{\theta}$ & $\unit[20]{mV}$ & Firing threshold \tabularnewline
$V_{\mathrm{reset}}$ & $\unit[0]{mV}$ & Reset potential \tabularnewline
$\tau_{\mathrm{ref}}$ & $\unit[2]{ms}$ & Absolute refractory period \tabularnewline
$\tau_{\mathrm{s}}$ & $\unit[0.5]{ms}$ & Postsynaptic current time constant \tabularnewline
\hline 
\end{tabular}

\begin{tabular}{|@{\hspace*{1mm}}p{0.1\linewidth}@{}|@{\hspace*{1mm}}p{0.2\linewidth}|@{\hspace*{1mm}}p{0.65\linewidth}|}
\hline 
\multicolumn{3}{|>{\color{white}\columncolor{black}}l|}{\textbf{C: Preprocessing}}\tabularnewline
\hline 
\textbf{Symbol} & \textbf{Value} & \textbf{Description}\tabularnewline
\hline 
$\Delta t$ & $\unit[1]{ms}$ & Temporal bin size\tabularnewline
$\Delta l$ & $\unit[0.1]{mm}$ & Spatial bin size\tabularnewline
\hline 
\end{tabular}

\caption{Simulation, network and preprocessing parameters. \label{tab:modelparams}}
\end{table}

\begin{table}[t]
\begin{tabular}{|@{\hspace*{1mm}}p{0.1\linewidth}@{}|@{\hspace*{1mm}}p{0.2\linewidth}|@{\hspace*{1mm}}p{0.65\linewidth}|}
\hline 
\multicolumn{3}{|>{\color{white}\columncolor{black}}l|}{\textbf{Simplified LFP model parameters}}\tabularnewline
\hline 
\textbf{Symbol} & \textbf{Value} & \textbf{Description}\tabularnewline
\hline 
$c_{\mathrm{m}}$ & $\unit[1]{\mu F/cm^{2}}$ & Specific membrane capacitance \tabularnewline
$r_{\mathrm{a}}$ & $\unit[150]{\Omega cm}$ & Axial resistivity \tabularnewline
$L_{\mathrm{dend}}$ & $\unit[500]{\mu m}$ & Dendritic stick length \tabularnewline
$r_{\mathrm{dend}}$ & $\unit[2.5]{\mu m}$ & Dendritic stick radius \tabularnewline
$n_{\mathrm{dend}}$ & \textbf{$11$} & Dendritic stick number of segments \tabularnewline
$r_{\mathrm{soma}}$ & $\unit[13.1]{\mu m}$ & Derived soma segment radius \tabularnewline
$\tau^{s}$ & $\unit[25]{ms}$ & Synapse activation time \tabularnewline
\selectlanguage{english}%
$\Delta l^{\phi}$\selectlanguage{american}%
 & $\unit[400]{\mu m}$ & Electrode separation, spatial bin width \tabularnewline
$\sigma_{\mathrm{e}}$ & $\unit[0.3]{S/m}$ & Extracellular conductivity \tabularnewline
\hline 
\end{tabular}

\caption{Parameters for prediction of LFP signals. \label{tab:LFP}}
\end{table}

\FloatBarrier

\end{document}